\documentclass[acmsmall]{acmart}
\usepackage{graphicx} %
\usepackage{enumitem}
\usepackage{appendix}
\usepackage{listings}  %
\usepackage{textcomp}
\usepackage{xcolor}
\usepackage{multirow}
\usepackage{colortbl}
\usepackage{subcaption}
\usepackage{enumitem}
\usepackage{algorithmic}
\usepackage{algorithm}
\usepackage{booktabs}
\usepackage{graphicx} 
\usepackage{bm}
\usepackage{xspace}
\usepackage{caption}
\usepackage{url}
\usepackage{tabularx}
\usepackage{courier}
\usepackage[normalem]{ulem}
\useunder{\uline}{\ul}{}
\usepackage[most]{tcolorbox}
\usepackage[figuresright]{rotating}
\usepackage{pdflscape}
\usepackage{pifont}

\usepackage{tablefootnote} %
\usepackage{tabularray}

\newcommand{\cl}[0]{\textsc{Code~Llama}}
\newcommand{\lm}[0]{\textsc{Llama~2}}
\newcommand{\cg}[0]{\textsc{CodeGen}}
\newcommand{\ic}[0]{\textsc{InCoder}}

\newcommand{\cc}[0]{\texttt{CCT5}}
\newcommand{\sm}[0]{small pre-trained model}

\newcommand{\icl}[0]{LLM-ICL}
\newcommand{\peft}[0]{LLM-PEFT}
\newcommand{\lora}[0]{LoRA}

\usepackage{todonotes}
\usepackage{hyperref}
\usepackage{marvosym}

\newcommand{\rqbox}[1]{

\begin{tcolorbox}[tile, size=fbox, boxsep=2mm, boxrule=0pt, top=0pt, bottom=0pt,
borderline west={1mm}{0pt}{blue!50!white}, colback=blue!5!white]
#1
\end{tcolorbox}

}

\renewcommand\footnotetextcopyrightpermission[1]{}

\definecolor{dkgreen}{rgb}{0,0.6,0}
\definecolor{gray}{rgb}{0.5,0.5,0.5}
\definecolor{mauve}{rgb}{0.58,0,0.82}
\title{Exploring the Capabilities of LLMs for Code Change Related Tasks}
\author{Lishui Fan}
\email{flscode@zju.edu.cn}
\authornote{Both authors contributed equally to the paper}
\affiliation{
    \institution{The State Key Laboratory of Blockchain and Data Security, Zhejiang University}
    \country{China}  
}
\author{Jiakun Liu}
\email{jkliu@smu.edu.sg}
\affiliation{
    \institution{Singapore Management University}
    \country{Singapore}  
}
\authornotemark[1]

\author{Zhongxin Liu}
\email{liu_zx@zju.edu.cn}
\authornote{Corresponding author}
\affiliation{
    \institution{The State Key Laboratory of Blockchain and Data Security, Zhejiang University}
    \country{China}  
}

\author{David Lo}
\email{davidlo@smu.edu.sg}
\affiliation{
    \institution{Singapore Management University}
    \country{Singapore}  
}

\author{Xin Xia}
\email{xin.xia@acm.org}
\affiliation{
    \institution{Zhejiang University}
    \country{China}  
}
 
\author{Shanping Li}
\email{shan@zju.edu.cn}
\affiliation{
    \institution{The State Key Laboratory of Blockchain and Data Security, Zhejiang University}
    \country{China}  
}

\date{September 2023}

\begin{abstract}

Developers deal with code-change-related tasks daily, e.g., reviewing code. Pre-trained code and code-change-oriented models have been adapted to help developers with such tasks. Recently, large language models (LLMs) have shown their effectiveness in code-related tasks. However, existing LLMs for code focus on general code syntax and semantics rather than the differences between two code versions. Thus, it is an open question how LLMs perform on code-change-related tasks.

To answer this question, we conduct an empirical study using \textgreater 1B parameters LLMs on three code-change-related tasks, i.e., code review generation, commit message generation, and just-in-time comment update, with in-context learning (ICL) and parameter-efficient fine-tuning (PEFT, including LoRA and prefix-tuning). We observe that the performance of LLMs is poor without examples and generally improves with examples, but more examples do not always lead to better performance. LLMs tuned with LoRA have comparable performance to the state-of-the-art small pre-trained models. Larger models are not always better, but \lm~and \cl~families are always the best. The best LLMs outperform small pre-trained models on the code changes that only modify comments and perform comparably on other code changes. We suggest future work should focus more on guiding LLMs to learn the knowledge specific to the changes related to code rather than comments for code-change-related tasks.

\end{abstract}

\begin{document}
\maketitle

\section{INTRODUCTION}
After the launch of a project, developers constantly change code to introduce new features and maintain existing code (e.g., performing refactoring and fixing bugs)~\cite{DBLP:conf/icse/Eilertsen20a,DBLP:conf/icse/DilharaDK23,DBLP:journals/tse/FluriWPG07}.
A code change contains the added, deleted, or modified (deleted then added) code span across one or more files and is often expressed in a combination of the code versions before and after the change, or in plain text such as ``diff''.
Developers need to handle many code-change-related tasks due to the ubiquitous nature of code changes.
For example, in their daily work, developers need to understand existing code changes in code repositories~\cite{fritz2010using}, update the comments accompanied with the changed code~\cite{DBLP:conf/paap/GuoXLC22}, write commit messages~\cite{DBLP:conf/kbse/LiuXHLXW18}, and review the code changed by other developers~\cite{DBLP:conf/icse/KovalenkoB18}.
These code-change-related tasks are important for project maintenance but can cost significant efforts and slow down the development process~\cite{bosu2013impact,lin1988classifying}.
Therefore, it is necessary to automate or provide tool support for them~\cite{lin1988classifying}.

To deal with code-change-related tasks, prior studies have proposed a series of machine-learning-based~\cite{kirinuki2014hey,kamei2016studying,rahman2016correct} and deep-learning-based approaches~\cite{lin2022predictive,liu2020automating,hoang2020cc2vec}.
Recently, researchers proposed to leverage pre-trained code models or pre-trained code-change-oriented models to tackle code-change-related tasks and achieved state-of-the-art performance~\cite{zhang2022coditt5,lin2023cct5,li2022automating}.
For example, \cc~\cite{lin2023cct5}, pre-trained based on CodeT5~\cite{wang2021codet5} using 1.5M code change samples with five code-change-oriented pre-training objectives, has achieved the state-of-the-art performance on diverse code-change-related tasks.
Recent studies have shown that by significantly increasing the model size and expanding the pre-training data, pre-trained models can be more powerful and can demonstrate various emergent abilities~\cite{wei2022emergent}.
For example, the Bloom model~\cite{muennighoff2022crosslingual} contains 176B parameters, 789 times more than \cc~(223M), is pre-trained on 363B natural language tokens, and significantly outperforms all the models with less than 1B parameters on natural language processing tasks.
The difference in model size and training data may make Bloom (or other similar models) perform better than small pre-trained models.

Recent work has pre-trained several \textgreater 1B parameters large language models (LLMs) with massive code corpora for code-related tasks~\cite{li2023starcoder,DBLP:journals/corr/abs-2306-08568,roziere2023code}.\footnote{For ease of explanation, we refer to \textgreater 1B parameters LLMs as simply LLMs, and those \textless 1B parameters LLMs as small pre-trained models.}
For example, Meta released the \cl~models~\cite{roziere2023code}, which are initialized from the \lm~models~\cite{touvron2023llama} and trained on 500B tokens from a code-heavy dataset.
However, these LLMs may not perform well for code-change-related tasks.
This can be the case as they are pre-trained with massive code snippets to learn the general syntactic and semantic knowledge of code,
while code changes are more about the differences between two code snippets.
Although we can represent a code change as a diff or other forms to help LLMs distinguish the changed parts from the context, LLMs are not pre-trained with the data of such formats.
Therefore, It is still an open question whether using \textgreater 1B parameters LLMs can effectively boost code-change-related tasks (as compared to smaller \textless 1B parameters LLMs).
To the best of our knowledge, there is a lack of an in-depth investigation of LLMs for code-change-related tasks.
In this work, we would like to fill this gap and investigate: \textbf{How do LLMs perform on code-change-related tasks?}

To answer this question, we select representative and popular \textgreater 1B parameters LLMs including \ic~\cite{fried2022incoder}, \cg~\cite{nijkamp2022codegen}, \lm~\cite{touvron2023llama}, and \cl~\cite{roziere2023code}.
We consider three emerging code-change-related tasks: code review generation \cite{fagan2011design}, commit message generation \cite{buse2010automatically}, and just-in-time comment update \cite{liu2020automating}.
We start the exploration of LLMs from prompt engineering.
Prompt engineering is the most convenient way to apply LLMs because it does not change model parameters.
In the realm of prompt engineering, In-Context Learning (ICL) is recognized as one of the most typical and effective approaches~\cite{geng2024large,li2023towards,meade2023using}.
It is a technique that formulates input by integrating task descriptions, exemplars, and query problems, and subsequently instructs LLMs to generate predictions.
To this end, we first would like to answer:\\
\noindent \textbf{RQ1: How do LLMs perform when applying In-Context Learning on code-change-related tasks?}\\
Hereon, we refer to LLMs with ICL as LLM-ICLs.
To answer this RQ, we apply LLM-ICLs with different numbers of examples on code-change-related tasks.
We find the performance of LLMs is poor without examples.
With one example provided, the performance of LLMs generally drastically improves.
However, more examples do not always lead to better performance.
The effectiveness of LLMs depends on the data lengths in the task and the context length allocated to the model.
Besides, larger models do not always have better performance, but models in \cl~family always perform the best in the selected tasks related to code changes.

To further explore the capabilities of LLMs on code-change-related tasks, we allow updating LLM parameters for code-change-related tasks.
Parameter-Efficient Fine-tuning Techniques (PEFT) are currently the most common technique for updating LLM parameters.
It focuses on updating a few parameters while freezing the rest, allowing the model to efficiently adapt to different tasks.
To this end, we summarize our second research question as:\\
\noindent \textbf{RQ2: How do LLMs perform when applying Parameter-Efficient Fine-tuning Techniques on code-change-related tasks?}\\
Hereon, we refer to LLMs tuned using PEFT as LLM-PEFTs.
We explore the capabilities of two most common and popular PEFT methods, i.e., LoRA~\cite{hu2021lora} and prefix-tuning~\cite{li2021prefix}.
LoRA can directly change the parameters of LLMs by optimizing the low-rank decomposition of LLMs' self-attention modules.
Prefix-tuning prepends a sequence of continuous trainable vectors to the input and the hidden states of each transformer layer.
We observe that LLMs tuned with \lora~results in significantly better performance compared to LLMs tuned using prefix-tuning.
Similarly, we also find that larger models do not necessarily have better performance, even within the same LLM family.
We also observe that the \lm~and \cl~families are the best-performing LLMs across the three code-change-related tasks.

Additionally, to help developers with the tasks related to code changes, previous researchers have proposed a series of tools based on \sm s.
We would like to further understand:\\
\noindent \textbf{RQ3: How do LLMs perform on code-change-related tasks compared to small pre-trained models?} \\
To answer this RQ, we compare the performance of LLMs to that of \sm s, i.e., CodeT5 \cite{wang2021codet5} (the small models pre-trained with code) and CCT5 \cite{lin2023cct5} (the small models pre-trained with code changes), on the selected code-change-related tasks.
We observe that LLM-ICLs are similar to or better than CodeT5 on the tasks related to code changes, but inferior to \cc~.
With parameter updates, LLM-PEFTs outperform LLM-ICLs across tasks.
Even though LLM-PEFTs have fewer parameters updated compared to the total number of parameters of small pre-trained models, LLM-PEFTs can still achieve comparable performance to \cc.

We notice that the code change can be presented in two different formats: in the diff format, or in two consecutive code snippets corresponding to the code before and after the change.
Intuitively, the input to an LLM may affect the performance of the LLM.
Therefore, we aim to explore:\\
\noindent \textbf{RQ4: How do LLMs perform with different input formats on code-change-related tasks?}\\
To answer this RQ, we compare the performance of LLMs with different input formats on the selected code-change-related tasks.
We observe that LLM-ICLs perform better when the input is in the ``diff'' format on the just-in-time comment update task.
For LLM-PEFTs, we find no significant difference among the three tasks with different input formats.
This may be because the updated parameters in the LLM have learned to compare the differences between two pieces of code and capture the patterns of the changed parts.

Finally, to further understand the better performance of LLM-PEFTs, we would like to explore:\\
\noindent \textbf{RQ5: When do LLMs perform better?} \\
We select the best-performing LLMs from the previous RQs and characterize their performance on different types of code changes.
We observe that LLM-PEFTs generally outperform LLM-ICLs on all types of code changes.
indicating that compared to ICL, PEFT can comprehensively improve the performance of LLMs on all code change categories.
LLM-PEFTs outperform the fully fine-tuned small models on the code changes that only revised documents in code (i.e., code comment), and on other code changes that perform code featuring code refactoring or modify both code and documentation, LLM-PEFTs perform comparably to fully fine-tuned small models.

We also conduct a human evaluation on commit message generation to further understand the effectiveness of LLMs.
The results show that the commit messages generated by LLM-PEFT are the most expressive and concise without losing much adequacy.

Our study illuminates the opportunities presented by LLMs, necessitating further investigations into their application in code-change-related tasks.
For example, we find LLMs tend to learn more knowledge related to documentation changes. We should focus more on guiding LLMs to learn the knowledge specific to the changes in code, such as the knowledge of refactoring, when tuning LLMs for code-change-related tasks.

In summary, this paper makes the following contributions:
\begin{itemize}[leftmargin=*]
	\item To the best of our knowledge, this paper presents the first comprehensive empirical study on the capabilities of LLMs in code-change-related tasks. We conduct experiments using a broad range of LLMs on three code-change-related tasks with different techniques and different input formats.

	\item Our comprehensive exploration and analysis highlight findings about how to apply LLMs to code-change-related tasks for different code-change types in different scenarios.

	\item  We discuss the implications of our findings and demonstrate the future work for code-change-related tasks in the era of LLMs.
\end{itemize}

\section{STUDY DESIGN}

\begin{figure*}[!ht]
    \centering
    \includegraphics[width=0.75\textwidth]{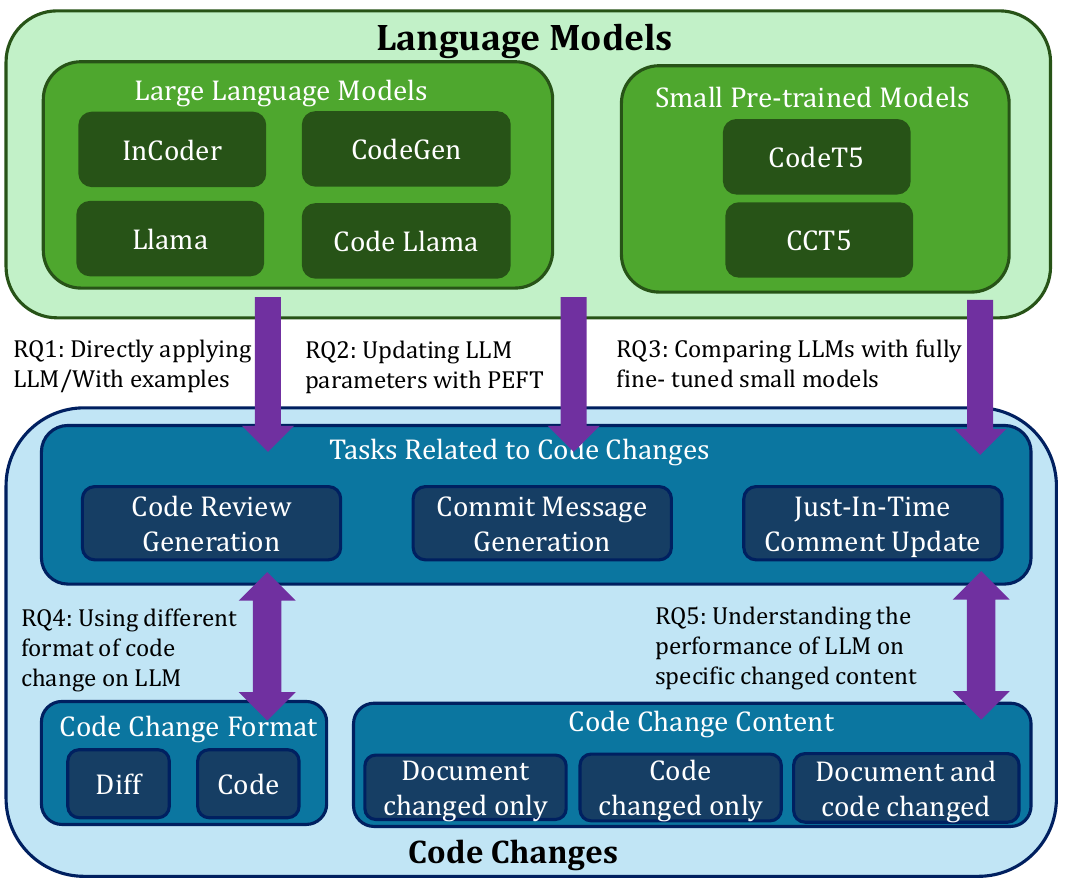}
    \caption{The overall framework of this study.}
    \label{fig_overview}
\end{figure*}

Figure \ref{fig_overview} presents the overview of our study.
We first present the selected code-changes-related tasks in Section \ref{sec_design_task}.
Then, in Section \ref{sec_design_model}, we present the state-of-the-art LLMs and \sm s.
We present the approaches we would like to explore to utilize these models (to answer RQ1, RQ2, and RQ3) in Section \ref{sec_design_technique}.
Following that, we present the design of input in Section \ref{sec_design_input} (to answer RQ4) and the design of analyzing the impact of code changes types in Section \ref{sec_design_content} (to answer RQ5).
Finally, we present the implementation of the work in Section \ref{sec_design_implementation}.

\subsection{Task, Dataset and Evaluation Metrics}\label{sec_design_task}

To perform our empirical study, we consider three emerging code-change-related tasks that have been popularly researched in recent years, namely, code review generation~\cite{li2022automating,siow2020core}, commit message generation ~\cite{dong2022fira,liu2020atom}, and just-in-time comment update ~\cite{zhu2022hatcup,panthaplackel2021deep}.

\subsubsection{Code Review Generation (CRG)}
The primary goal of CRG is to automatically generate reviewers' suggestions from code changes to provide immediate high-quality feedback to developers when they commit to the version control system.
This task takes as input a code change included in the commit and generates the corresponding code review comment as output.

\noindent\textbf{Datasets.} Following prior studies~\cite{lin2023cct5,liu2022commitbart}, we use the dataset for the CRG task constructed by Li et al.~\cite{li2022automating}. They collected a total of 138,127 diff-review pairs from popular open-source projects on GitHub written in nine different programming languages. Each pair consists of a real-world code change along with the corresponding code review comment.

\noindent\textbf{Metrics.} To evaluate the generated text, following Li et al.\cite{li2022automating}'s work, we utilize the BLEU-4 metric. BLEU-4 computes the overlap of n-grams between the generated and the reference texts, where n ranges from 1 to 4.

\vspace{-3pt}
\subsubsection{Commit Message Generation (CMG)} The primary goal of CMG is to automatically produce a concise natural language description to summarize the content and intention of a commit submitted to the version control system. The task takes as input the code change in the commit and generates the corresponding code commit message as output.

\noindent\textbf{Dataset.} Following prior studies \cite{lin2023cct5,shi2022race}, we use the Multi-programming language Commit Message Dataset (MCMD) constructed by Tao et al.~\cite{tao2022large}.
They considered 5 programming languages, i.e., Python, Java, JavaScript, C\#, and C++, and collected the top 100 starred projects in each programming language from GitHub respectively (500 projects in total).\footnote{In this paper, we refer to them as CMG-Python, CMG-Java, CMG-JavaScript, CMG-C\#, and CMG-C++ respectively.}
Then, for each programming language, they randomly sampled 450,000 commits as well as their corresponding commit messages from the collected projects (2,250,000 commits in total).
Note that the original MCMD dataset only provides the tokenized source code data. For example, the code snippet \textit{this.indices.limit(indices.length);} has been tokenized into \textit{this . indices . limit ( indices . length ) ;}.
Considering that the input format can impact the performance of LLMs and different models may use different tokenizers, we used regular expressions to restore the data in MCMD by removing additional spaces.
For example, \textit{this . indices . limit ( indices . length ) ;} will be converted back to \textit{this.indices.limit(indices.length);}. Each model will be provided with the de-tokenized data and will use its own tokenizer. It is worth noting that we only changed the format of the input, while the content of the input remained unchanged.

\noindent\textbf{Metrics.} Following prior studies~\cite{tao2022large,loyola2017neural}, we use B-Norm as the evaluation metric. B-Norm is a variant of BLEU \cite{papineni2002bleu}, which assesses the lexical overlap between the generated text and the label. Prior work has shown that B-Norm demonstrates the highest correlation with human evaluations on CMG~\cite{tao2022large}. For convenience, on the CMG task, we also refer to B-Norm as BLEU.

\vspace{-3pt}
\subsubsection{Just-In-Time Comment Update (JITCU)} The primary goal of just-in-time (JIT) comment update is to automatically update comments after code changes, which can avert outdated comments and boost the maintainability of software. This task takes as input a code change and the comment associated with the modified code in the before-change version and generates the updated comments after the change.

\noindent\textbf{Dataset.} Following prior studies\cite{DBLP:conf/iwpc/LinW0MB21,lin2023cct5}, we use the dataset that was first constructed by Liu et al. ~\cite{DBLP:conf/kbse/LiuXYL20} and then further cleaned by Lin et al. ~\cite{DBLP:conf/iwpc/LinW0MB21} for JITCU. This dataset contains a total of 98,622 comment update instances collected from 1,496 high-quality Java projects on GitHub. Each data entry comprises co-changes between methods and header comments.

\noindent\textbf{Metrics.} Following prior studies~\cite{DBLP:journals/tse/LiuXLYL23,zhang2022coditt5,lin2023cct5}, we use GLUE and ACC as the evaluation metrics. GLEU is similar to BLEU, and is widely used to evaluate text-editing systems. Additionally, we also use Accuracy (shortened as ACC), which is computed as the percentage of the test instances where the generated comments are the same as the ground truth.

Following prior studies~\cite{koehn2004statistical,liu2020automating}, we also use paired bootstrap resampling with 1000 resamples
\cite{koehn2004statistical} to perform statistical significance tests for each evaluation metric.

\subsection{Small Pre-Trained Models and Large Language Models}\label{sec_design_model}
\label{sec:models}
In this study, we consider the pre-trained language models with fewer than 1B parameters as small pre-trained models and those with more than 1B parameters as large language models.

\textbf{To select small pre-trained models}, we select the state-of-the-art models on code-related tasks and code-change-related tasks, namely CodeT5~\cite{wang2021codet5} and CCT5~\cite{lin2023cct5}, in our experiments.

\textbf{CodeT5} is an encoder-decoder model. It is pre-trained with the code data from CodeSearchNet~\cite{husain2019codesearchnet} and the C and CSharp code data from BigQuery\footnote{https://console.cloud.google.com/ marketplace/details/github/github-repos} using four pre-training tasks, such as Identifier Tagging and Masked Identifier Prediction. Existing code-change-oriented pre-trained models~\cite{lin2023cct5,zhang2022coditt5,li2022automating} are all built upon CodeT5. Following prior studies~\cite{ayupov2022parameter,zhang2022coditt5,li2022automating}, we use the base model with 223M parameters.

\textbf{\cc} is the state-of-the-art code-change-oriented pre-trained models. It is pre-trained with 39.6 GB of code change data collected from 35K popular GitHub repositories in six programming languages. Five code-change-oriented tasks are used for pre-training, including masked language modeling for code change, code diff generation, code diff to natural language generation, etc.

\textbf{To select LLMs,} following prior work~\cite{weyssow2023exploring}, we use the following criteria:

\begin{itemize}[leftmargin=*]

    \item We select open-source LLMs and exclude closed-source LLMs. Because the parameters of closed-source LLMs are inaccessible, making the investigation of fine-tuning techniques unfeasible or expensive.

    \item We select state-of-the-art LLMs in the software engineering field, especially those released recently at the time we conducted this study (i.e., September 2023)..

    \item We select a model family in the general domain, facilitating the comparison between the LLMs in the code and general domains

    \item We select models across different parameter sizes to study the implications of parameter sizes.
\end{itemize}

Consequently, we select four LLM families, \cg ~\cite{nijkamp2022codegen}, \ic ~\cite{fried2022incoder}, \cl~\cite{roziere2023code} and \lm~\cite{touvron2023llama}.

\begin{table}[!htbp]
  \caption{Small pre-trained models and LLM families included in our study. For the models reported with two parameter sizes, we use both of them in our work.}
  \label{table_selected_llms}
  \centering
  \resizebox{0.7\linewidth}{!}{
  \begin{tabular}{cl|ccc}
    \hline
    \multicolumn{2}{c|}{Models}                                                & Dataset    & Parameters    & Domain               \\ \hline
    \multirow{2}{*}{\begin{tabular}[c]{@{}c@{}}Small\\ PT Models\end{tabular}} &
    CodeT5                                                                     &
    CodeSearchNet                                                              &
    223M                                                                       &
    Code                                                                                                                           \\
                                                                               & \cc       & CodeChangeNet & 223M   & Code Change \\ \hline
    \multirow{6}{*}{LLMs}                                                      & \ic    & -             & 1.3B   & Code        \\
                                                                               &
    \cg                                                                    &
    \begin{tabular}[c]{@{}c@{}}The Pile /\\ BigQuery \end{tabular}             &
    2B/6B                                                                      &
    Code                                                                                                                           \\
                                                                               & \lm    & -             & 7B/13B & General     \\
                                                                               & \cl & -             & 7B/13B & Code        \\ \hline
  \end{tabular}
  }
\end{table}

\textbf{\ic} models are built upon XGLM~\cite{lin2021few} and is pre-trained using the infilling task.
We utilize \ic-1.3B in our experiments.

\textbf{\cg} models are built upon GPT-Neo~\cite{gpt-neo} and GPT-J~\cite{gpt-j}, and is pre-trained using the autoregressive language modeling task. In our experiments, we choose \cg-Multi-2B and \cg-Multi-6B, which are pre-trained on the code data in a wide range of programming languages collected from BigQuery.

\textbf{\lm} is pre-trained on 2 trillion tokens of data using the autoregressive language modeling task. In our experiments, considering the computation cost, we employ both \lm-7B and \lm-13B in our experiments.

\textbf{\cl} family are based upon \lm. It is pre-trained on a 500B token corpus with both code and natural language texts using the infilling objective. Similar to \lm, in our experiments, we employ both \cl-7B and \cl-13B in our experiments.
We exclude the variants that are further trained on Python corpus (too specific) or instruction database (different capability) as they are not suitable for our task.

\subsection{Techniques to apply LLMs}\label{sec_design_technique}
To understand the capability of LLMs on code-change-related tasks, we first explore prompt engineering.
Because prompt engineering does not require updating model parameters and is convenient and cheap.
Then, we explore the techniques of changing parts of model parameters, namely parameter-efficient fine-tuning (PEFT) techniques.
In addition, we also explore the method of changing all model parameters, namely full fine-tuning.

\noindent \textbf{Prompt Engineering} design proper prompts to guide the pre-trained model to perform a specific downstream task. We choose the most popular technique, i.e., in-context learning (ICL), in prompt engineering. ICL is a technique that guides LLMs to generate contextually appropriate content without updating parameters by providing examples or demonstrations of the task in the prompt.
Prior studies have demonstrated the effectiveness of LLM-ICLs on domain-specific tasks \cite{geng2024large,li2023towards,meade2023using}.

We explore the capability of ICL on code-change-related tasks by varying the number of similar examples, denoted as $n$, where $n \in {0, 1, 2, 3, 4}$.
$n = 0$ corresponds to the zero-shot scenario~\cite{brown2020language}, enabling us to explore the capability of LLMs without any prior knowledge of the task.
For each test sample, we employ the BM25 method to retrieve similar samples from the training set as examples. BM25, a well-established information retrieval technique, has been shown to perform well to retrieve demonstrations for ICL by previous studies~\cite{gao2023constructing,meade2023using}.
Additionally, we do not explore the performance of ICL with small pre-trained models, because it is hard for small pre-trained models to comply with instructional prompts without parameter updates~\cite{mishra2021cross,raffel2020exploring}.

\noindent \textbf{Parameter-efficient fine-tuning} (PEFT) adapts LLMs to downstream tasks by updating a minimal subset of model parameters, either inherent or newly added to the model, with task-specific datasets.
We choose two most popular PEFT techniques: LoRA and prefix-tuning.

\begin{itemize}
    \item \textbf{\textit{LoRA}} is proposed to update partial parameters of LLMs by optimizing the low-rank decomposition of the attention module's matrices. It is widely used on domain-specific tasks and has proven to be effective~\cite{lu2023llama,liu2023radiology,ye2023qilin}. When using LoRA, we need to configure two hyper-parameters $\gamma$ and $\alpha$, where $\gamma$ is the rank of the update matrices and $\alpha$ is a scaling factor that helps stabilize the training. Following prior work~\cite{hu2021lora}, we consider $\gamma=8$ and $\alpha=16$.

    \item \textbf{\textit{Prefix-tuning}} wraps the input with additional context by prepending trainable continuous vectors (prefixes) to the input and the hidden states of each transformer layer. It is broadly used on domain-specific tasks and has established its effectiveness~\cite{chen2023one,zhao2022domain}. When using prefix-tuning, we need to configure the number of trainable vectors $n$. Following previous work~\cite{li2021prefix}, we set $n=8$.

\end{itemize}
\noindent \textbf{Full-Parameter Fine-tuning.} This technique updates all the parameters of a model for one or more tasks and often consumes a large amount of resources. Due to computational resource limitations, we are only able to fine-tune the small pre-trained models, i.e., CodeT5 and \cc, using full-parameter fine-tuning.

\begin{table}[t]
  \caption{An overview of the input design in code change tasks under two different input formats. The \textit{\{input\_tokens\}} are different in the diff input format and code input format.}
  \label{table_input}
  \centering
  \resizebox{0.6\linewidth}{!}{
  \begin{tabular}{l|l}
    \hline
    Tasks &
    \multicolumn{1}{c}{Input Design(diff/code)}                                                                                                                                                                                   \\ \hline
    CRG   &
    \begin{tabular}[c]{@{}l@{}}\#\#\# Instruction:\\ Please write a code review according to the \\ (diff hunk/code before and after the diff hunk).\\ \{input\_tokens\}\\ \#\#\# Answer:\end{tabular}                            \\ \hline
    CMG   &
    \begin{tabular}[c]{@{}l@{}}\#\#\# Instruction:\\ Please write a commit message according to the\\ (diff hunk/code before and after the diff hunk).\\ \{input\_tokens\}\\ \#\#\# Answer:\end{tabular}                          \\ \hline
    JITCU &
    \begin{tabular}[c]{@{}l@{}}\#\#\# Instruction:\\ Please write a new comment according to the\\ original comment and the (diff hunk/code before \\ and after the diff hunk).\\ \{input\_tokens\}\\ \#\#\# Answer:\end{tabular} \\ \hline
  \end{tabular}
  }
\end{table}

\subsection{Input Design}\label{sec_design_input}

Table \ref{table_input} shows the prompt templates for the selected tasks.
Each template contains ``\#\#\# Instruction:", the description of the task, examples, and the code change needed to process.
Finally, we use ``\#\#\# Answer:" to introduce the response, such as a commit message, code comment, or code review comment.
Specifically, each example is composed of a code change and the corresponding expected response (i.e., the ground truth appended after ``\#\#\# Answer:").
Additionally, for the JITCU task, where each input contains an original comment, we add ``original comment message:$\backslash$n" to the end of the code change.
For the techniques that do not need examples (LLM-ICL without example and PEFT), there is no example in the prompt.
Note that while the format of these prompt templates may not be optimal, its effectiveness has been demonstrated in prior studies~\cite {peng2023instruction,pryzant2023automatic}.

To investigate the impact of different formats of code change on the model performance, we explore the code changes presented in the diff format or in the code format.
Specifically, the diff format highlights the lines of code that have been added or removed with ``+'' or ``-'' at the beginning of the line, respectively; while the code format puts code snippets before and after the change together.
To use them in LLM, for diff format, we add the ``diff hunk:$\backslash$n" before the diff text;
for code format, we add ``code change before:$\backslash$n" and ``code change after:$\backslash$n" before the code snippet before and after the change, respectively, and then connect the two parts with ``$\backslash$n".

\subsection{Analyzing the Impact of Code Change Types}\label{sec_design_content}

To understand the performance of different models on different types of code changes, for each code-change-related task, we select the best-performing models from small pre-trained models, LLM-ICLs, and LLM-PEFTs.
Then we randomly selected 592 samples from the test datasets with a 90\% confidence level and a 5.6\% confidence interval.
The selected dataset contains 196 samples from CRG, 196 samples from JITCU, and 200 samples from CMG (40 for each language).
Two authors manually annotated the change category of each sample independently.
We use the taxonomy of code changes proposed in prior studies \cite{guo2023exploring,tufano2023automating} as a starting point and refine the taxonomy based on our observation.
As a result, we consider three categories of code changes: \textit{Doc-only code change, Code-only code change, Doc-and-Code code change}.

\begin{itemize}[wide=0pt]
    \item \textbf{Doc-only code change} represents the code changes that only add, modify or remove documentation.
    \item \textbf{Code-only code change} represents the code changes which only modify code entities. Furthermore, this category can be divided into two subcategories: \ding{172} \textbf{Feature code change} represents the code-only code changes where the functional logic is modified.
          \ding{173} \textbf{Refactor code change} refers to the code-only code changes that perform code refactoring, including renaming code entities, swapping two code snippets, and updating code based on coding standards.
    \item \textbf{Doc-and-Code code change} represents the code changes that include both documentation and code modifications.
\end{itemize}

After the two authors independently annotated the selected samples, a discussion was held to solve the disagreements.
We did not invite others because all the disagreements were resolved during the discussion.
Finally, we report the performance of different models on different types of code changes.

\subsection{Implementation}
\label{sec_design_implementation}
Our implementation is based on the Huggingface\footnote{https://huggingface.co/} library.
Specifically, we use this library to download the models and their tokenizers and to conduct all experiments in our paper. Note that we do not apply prefix-tuning to the \ic~model because there is an implementation issue with adding the virtual tokens of prefix-tuning to the base model of \ic~in the Huggingface library.
Such an issue has also been mentioned by other developers\footnote{https://github.com/huggingface/peft/issues/811} and researchers ~\cite{weyssow2023exploring}.

Following prior studies~\cite{chai2022ernie,weyssow2023exploring}, we use half-precision for LLMs to fit them into our GPU, and full precision for the small pre-trained models to ensure their performance.
To make a fair comparison between models, following the prior studies \cite{gao2023constructing,yuan2023evaluating,raschka2018model,yadav2016analysis,zeid2022efficient} that considered the non-trivial costs of fine-tuning LLMs, we sample 16,000, 2,000 and 2,000 samples from the original dataset (or each sub-dataset in MCMD \cite{tao2022large}) as the training, validation, and test sets for each task.
Note that prior studies \cite{lin2023cct5, shi2022race} split training, validation, and test sets from the whole dataset to fully explore the performance of the small model; considering the different sizes of the training data, our experimental results can be different from theirs \cite{lin2023cct5, shi2022race}.

For each task, we use a maximum length of 1024 for the input of each sample.
Because we find the input lengths of over 98\% of the samples in the training sets of the three tasks are less than 1024.
Particularly, we allow ICL to use another 1024 tokens to provide examples in the prompt.
In detail, when the number of examples is $n$, the maximum length for each example is set to $\frac{1024}{n}$ (n \textgreater 0).
When the input length exceeds the maximum length, we proportionally truncate the code snippets before and after the change at the same time or truncate the diff, depending on the input format.
For JITCU, if the input length exceeds the maximum length, we prioritize truncating the code or diff while preserving the integrity of the original comments.
For output, we consider a maximum length of 100 tokens.
This is because for over 97\% of the examples in the training sets of the three tasks, their reference output has less than 100 tokens.

\section{Experimental Results}

Here, we present the results of the experiments to answer the five research questions.

\subsection{RQ1: How Do LLMs Perform When Applying In-Context Learning on Code-Change-Related Tasks?}\label{rq_icl}

\begin{table*}[!htbp]
\centering
    \caption{ {[}RQ1{]} - BLEU scores of LLM-ICLs on CRG. The darker color of the cells means better performance.}
    \label{table_icl_crg}
    \resizebox{0.9\textwidth}{!}{
    \begin{tabular}{lccccccccc}
        \toprule
        Models    & 0shot                          & 1shot                          & 2shot                                 & 3shot                          & 4shot                          & 5shot                          & 6shot                          & 7shot                                  & 8shot                                 \\ \midrule
        \ic-1b    & \cellcolor[HTML]{FFD5B6}{1.44} & \cellcolor[HTML]{FFA35C}{3.22} & \cellcolor[HTML]{FF9B4D}{\uline{3.5}} & \cellcolor[HTML]{FF9B4F}{3.47} & \cellcolor[HTML]{FFA35D}{3.19} & \cellcolor[HTML]{FFA966}{3.01} & \cellcolor[HTML]{FFAC6D}{2.88} & \cellcolor[HTML]{FFAF72}{2.78}         & \cellcolor[HTML]{FFAF71}{2.8}         \\
        \cg-2b-nl    & \cellcolor[HTML]{FFFAF6}{0.16} & \cellcolor[HTML]{FFF8F3}{0.22} & \cellcolor[HTML]{FFE3CE}{0.95}        & \cellcolor[HTML]{FFDBC0}{1.23} & \cellcolor[HTML]{FFD7B9}{1.38} & \cellcolor[HTML]{FFD5B5}{1.45} & \cellcolor[HTML]{FFD3B2}{1.51} & \cellcolor[HTML]{FFD1AF}{1.58}         & \cellcolor[HTML]{FFD1AE}{\uline{1.6}} \\
        \cg-6b-nl & \cellcolor[HTML]{FFEFE2}{0.56} & \cellcolor[HTML]{FFE9D8}{0.76} & \cellcolor[HTML]{FFE4CF}{0.93}        & \cellcolor[HTML]{FFDDC3}{1.18} & \cellcolor[HTML]{FFD8BB}{1.34} & \cellcolor[HTML]{FFD8BB}{1.34} & \cellcolor[HTML]{FFD1AE}{1.6}  & \cellcolor[HTML]{FFD1AE}{1.6}          & \cellcolor[HTML]{FFC599}{\uline{2.0}} \\
        \lm-7b    & \cellcolor[HTML]{FFECDE}{0.65} & \cellcolor[HTML]{FF8120}{4.4}  & \cellcolor[HTML]{FF8120}{4.4}2        & \cellcolor[HTML]{FF7D18}{4.55} & \cellcolor[HTML]{FF7E1B}{4.49} & \cellcolor[HTML]{FF7810}{4.72} & \cellcolor[HTML]{FF760C}{4.79} & \cellcolor[HTML]{FF7206}{\uline{4.91}} & \cellcolor[HTML]{FF7308}{4.88}        \\
        \lm-13b   & \cellcolor[HTML]{FFCDA7}{1.72} & \cellcolor[HTML]{FF8527}{4.25} & \cellcolor[HTML]{FF8120}{4.4}2        & \cellcolor[HTML]{FF7B15}{4.62} & \cellcolor[HTML]{FF7912}{4.68} & \cellcolor[HTML]{FF7810}{4.72} & \cellcolor[HTML]{FF770F}{4.73} & \cellcolor[HTML]{FF7307}{4.89}         & \cellcolor[HTML]{FF7002}{\uline{5.0}} \\
        \cl-7b    & \cellcolor[HTML]{FFE5D1}{0.9}  & \cellcolor[HTML]{FF7D18}{4.55} & \cellcolor[HTML]{FF770F}{4.74}        & \cellcolor[HTML]{FF750A}{4.83} & \cellcolor[HTML]{FF7002}{5}.02 & \cellcolor[HTML]{FF7103}{4.97} & \cellcolor[HTML]{FF7002}{5}.03 & \cellcolor[HTML]{FF7002}{\uline{5.04}} & \cellcolor[HTML]{FF7205}{4.93}        \\
        \cl-13b   & \cellcolor[HTML]{FFC599}{2}.27 & \cellcolor[HTML]{FF7B16}{4.6}  & \cellcolor[HTML]{FF7B16}{4.6}5        & \cellcolor[HTML]{FF760C}{4.79} & \cellcolor[HTML]{FF750C}{4.8}  & \cellcolor[HTML]{FF750C}{4.8}6 & \cellcolor[HTML]{FF760C}{4.79} & \cellcolor[HTML]{FF7104}{\uline{4.95}} & \cellcolor[HTML]{FF7105}{4.94}        \\ \bottomrule
    \end{tabular}
     }
\end{table*}

\begin{table*}[!htbp]
\centering
    \caption{{[}RQ1{]} - BLEU scores of LLM-ICLs on Python and Java sub-datasets of CMG. The darker color of the cells means better performance.}
    \label{table_icl_cmg}
    \resizebox{\textwidth}{!}{
    \begin{tabular}{clccccccccc}
        \toprule
        \multicolumn{1}{l}{Lang} & Model     & 0shot                                  & 1shot                                  & 2shot                                  & 3shot                                  & 4shot                                 & 5shot                                  & 6shot                          & 7shot                                  & 8shot                                  \\ \midrule
                                 & \ic-1b    & \cellcolor[HTML]{FFC191}{\uline{3.59}} & \cellcolor[HTML]{FFCCA4}{2.97}         & \cellcolor[HTML]{FFCEA9}{2.8}          & \cellcolor[HTML]{FFCDA6}{2.91}         & \cellcolor[HTML]{FFD2B0}{2.58}        & \cellcolor[HTML]{FFCEA9}{2.8}6         & \cellcolor[HTML]{FFCFAA}{2.79} & \cellcolor[HTML]{FFCBA3}{3}            & \cellcolor[HTML]{FFD0AC}{2.71}         \\
                                 & \cg-2b-nl & \cellcolor[HTML]{FFDEC5}{1.9}          & \cellcolor[HTML]{FFD9BB}{2.21}         & \cellcolor[HTML]{FFD5B5}{2.41}         & \cellcolor[HTML]{FFD3B1}{\uline{2.54}} & \cellcolor[HTML]{FFD5B5}{2.41}        & \cellcolor[HTML]{FFD6B7}{2.34}         & \cellcolor[HTML]{FFD9BB}{2.21} & \cellcolor[HTML]{FFDBBF}{2.09}         & \cellcolor[HTML]{FFD8BA}{2.26}         \\
                                 & \cg-6b-nl & \cellcolor[HTML]{FFDABF}{2.1}          & \cellcolor[HTML]{FFCEA9}{2.8}7         & \cellcolor[HTML]{FFCBA3}{\uline{3.28}} & \cellcolor[HTML]{FFCBA3}{3}.17         & \cellcolor[HTML]{FFCBA3}{3}.13        & \cellcolor[HTML]{FFCBA3}{3}.07         & \cellcolor[HTML]{FFCBA3}{3}.09 & \cellcolor[HTML]{FFCBA3}{3}.16         & \cellcolor[HTML]{FFCBA3}{3}.11         \\
                                 & \lm-7b    & \cellcolor[HTML]{FFD8BB}{2.23}         & \cellcolor[HTML]{FFCBA3}{3}.48         & \cellcolor[HTML]{FFCBA3}{3}.63         & \cellcolor[HTML]{FFCBA3}{3}.95         & \cellcolor[HTML]{FFCBA3}{3}.97        & \cellcolor[HTML]{FFB882}{4.08}         & \cellcolor[HTML]{FFB780}{4.16} & \cellcolor[HTML]{FFB984}{4.04}         & \cellcolor[HTML]{FFB77F}{\uline{4.18}} \\
                                 & \lm-13b   & \cellcolor[HTML]{FFD5B5}{2.42}         & \cellcolor[HTML]{FFCBA3}{3}.68         & \cellcolor[HTML]{FFCBA3}{3}.8          & \cellcolor[HTML]{FFB073}{4.58}         & \cellcolor[HTML]{FFAB6B}{4.84}        & \cellcolor[HTML]{FFA25B}{5.39}         & \cellcolor[HTML]{FF9C50}{5.75} & \cellcolor[HTML]{FF9E53}{5.65}         & \cellcolor[HTML]{FF9848}{\uline{6.00}} \\
                                 & \cl-7b    & \cellcolor[HTML]{FFE0C8}{1.78}         & \cellcolor[HTML]{FF9F56}{5.55}         & \cellcolor[HTML]{FFA45E}{5.29}         & \cellcolor[HTML]{FFA35D}{5.32}         & \cellcolor[HTML]{FFA25B}{5.38}        & \cellcolor[HTML]{FFA25B}{5.37}         & \cellcolor[HTML]{FF9848}{6}.07 & \cellcolor[HTML]{FF9848}{\uline{6.19}} & \cellcolor[HTML]{FF9848}{6}.12         \\
        \multirow{-7}{*}{Java}   & \cl-13b   & \cellcolor[HTML]{FFD1AD}{2.68}         & \cellcolor[HTML]{FFB278}{\uline{4.44}} & \cellcolor[HTML]{FFCBA3}{3}.76         & \cellcolor[HTML]{FFCBA3}{3}.97         & \cellcolor[HTML]{FFBA85}{4.01}        & \cellcolor[HTML]{FFB881}{4.12}         & \cellcolor[HTML]{FFB57C}{4.28} & \cellcolor[HTML]{FFA967}{4.98}         & \cellcolor[HTML]{FFAD6E}{4.74}         \\ \midrule
                                 & \ic-1b    & \cellcolor[HTML]{FFA763}{\uline{5.11}} & \cellcolor[HTML]{FFCBA3}{3}.97         & \cellcolor[HTML]{FFB881}{4.12}         & \cellcolor[HTML]{FFB57C}{4.28}         & \cellcolor[HTML]{FFB47B}{4.32}        & \cellcolor[HTML]{FFB57C}{4.3}          & \cellcolor[HTML]{FFB276}{4.48} & \cellcolor[HTML]{FFCBA3}{3}.98         & \cellcolor[HTML]{FFB77F}{4.19}         \\
                                 & \cg-2b-nl & \cellcolor[HTML]{FFD2B0}{2.58}         & \cellcolor[HTML]{FFCBA3}{3}.23         & \cellcolor[HTML]{FFCBA3}{3}.24         & \cellcolor[HTML]{FFCBA3}{3}.28         & \cellcolor[HTML]{FFCBA3}{3}.26        & \cellcolor[HTML]{FFCBA3}{3}.27         & \cellcolor[HTML]{FFCBA3}{3}.21 & \cellcolor[HTML]{FFCBA3}{3}.24         & \cellcolor[HTML]{FFCBA3}{\uline{3.44}} \\
                                 & \cg-6b-nl & \cellcolor[HTML]{FFCBA3}{3}.4          & \cellcolor[HTML]{FFB77F}{4.19}         & \cellcolor[HTML]{FFAE6F}{\uline{4.71}} & \cellcolor[HTML]{FFB278}{4.43}         & \cellcolor[HTML]{FFB57C}{4.3}4        & \cellcolor[HTML]{FFB77F}{4.19}         & \cellcolor[HTML]{FFB67E}{4.24} & \cellcolor[HTML]{FFB57C}{4.3}5         & \cellcolor[HTML]{FFB57C}{4.3}5         \\
                                 & \lm-7b    & \cellcolor[HTML]{FFCBA3}{3}.53         & \cellcolor[HTML]{FFB378}{4.41}         & \cellcolor[HTML]{FFB073}{4.59}         & \cellcolor[HTML]{FF9F55}{5.58}         & \cellcolor[HTML]{FF9848}{6}.19        & \cellcolor[HTML]{FF9848}{\uline{6.37}} & \cellcolor[HTML]{FF9848}{6}.3  & \cellcolor[HTML]{FF9848}{6}.05         & \cellcolor[HTML]{FF9848}{6}.33         \\
                                 & \lm-13b   & \cellcolor[HTML]{FFB378}{4.41}         & \cellcolor[HTML]{FFA763}{5.12}         & \cellcolor[HTML]{FF9848}{6}.15         & \cellcolor[HTML]{FF9848}{6}.75         & \cellcolor[HTML]{FF8325}{7.17}        & \cellcolor[HTML]{FF801F}{7.36}         & \cellcolor[HTML]{FF7D1A}{7.52} & \cellcolor[HTML]{FF8221}{7.28}         & \cellcolor[HTML]{FF7C17}{\uline{7.63}} \\
                                 & \cl-7b    & \cellcolor[HTML]{FFE5D1}{1.51}         & \cellcolor[HTML]{FF9848}{6}.19         & \cellcolor[HTML]{FF8221}{7.28}         & \cellcolor[HTML]{FF7912}{7.79}         & \cellcolor[HTML]{FF6F00}{8.36}        & \cellcolor[HTML]{FF6F00}{\uline{8.39}} & \cellcolor[HTML]{FF7307}{8.13} & \cellcolor[HTML]{FF7206}{8.18}         & \cellcolor[HTML]{FF7409}{8.07}         \\
        \multirow{-7}{*}{Python} & \cl-13b   & \cellcolor[HTML]{FFCBA3}{3}.63         & \cellcolor[HTML]{FFA45F}{5.26}         & \cellcolor[HTML]{FF9848}{6}.42         & \cellcolor[HTML]{FF8221}{7.28}         & \cellcolor[HTML]{FF7A14}{\uline{7.7}} & \cellcolor[HTML]{FF7C18}{7.59}         & \cellcolor[HTML]{FF9848}{6}.95 & \cellcolor[HTML]{FF8121}{7.3}          & \cellcolor[HTML]{FF9848}{6}.29         \\ \bottomrule
    \end{tabular}
    }
\end{table*}

\begin{table}[htbp]
    \caption{{[}RQ1{]} - Performance of LLM-ICLs on JITCU. We report the GLEU and ACC. The darker color of the cells means better performance.}
    \label{table_icl_jitcu}
    \centering
    \resizebox{\textwidth}{!}{
    \begin{tabular}{lcccccccccc}
        \toprule
        Model     & \multicolumn{1}{l}{metric} & 0shot                          & 1shot                           & 2shot                                   & 3shot                                   & 4shot                           & 5shot                                  & 6shot                                  & 7shot                                   & 8shot                           \\ \midrule
        \ic-1b    &                            & \cellcolor[HTML]{FFEFE3}{6.54} & \cellcolor[HTML]{FFE2CC}{11.94} & \cellcolor[HTML]{FFCAA1}{21.92}         & \cellcolor[HTML]{FFB983}{\uline{28.93}} & \cellcolor[HTML]{FFBC89}{27.55} & \cellcolor[HTML]{FFBB87}{28.13}        & \cellcolor[HTML]{FFBF8F}{26.27}        & \cellcolor[HTML]{FFBF8E}{26.47}         & \cellcolor[HTML]{FFD2B0}{18.41} \\
        \cg-2b-nl &                            & \cellcolor[HTML]{FFFFFF}{0.00} & \cellcolor[HTML]{FFFEFE}{0.06}  & \cellcolor[HTML]{FFFCF9}{1.18}          & \cellcolor[HTML]{FFFCF9}{1.20}          & \cellcolor[HTML]{FFFBF8}{1.44}  & \cellcolor[HTML]{FFFAF7}{1.86}         & \cellcolor[HTML]{FFF9F5}{\uline{2.30}} & \cellcolor[HTML]{FFFAF6}{2.01}          & \cellcolor[HTML]{FFFAF6}{1.95}  \\
        \cg-6b-nl &                            & \cellcolor[HTML]{FFFFFF}{0.00} & \cellcolor[HTML]{FFF8F3}{2.68}  & \cellcolor[HTML]{FFF1E6}{5.68}          & \cellcolor[HTML]{FFDFC6}{13.18}         & \cellcolor[HTML]{FFDFC6}{13.26} & \cellcolor[HTML]{FFD8BB}{15.85}        & \cellcolor[HTML]{FFD7B9}{16.33}        & \cellcolor[HTML]{FFD5B5}{\uline{17.36}} & \cellcolor[HTML]{FFD5B6}{17.05} \\
        \lm-7b    &                            & \cellcolor[HTML]{FFF3EA}{4.90} & \cellcolor[HTML]{FF8323}{51.49} & \cellcolor[HTML]{FF7912}{\uline{55.55}} & \cellcolor[HTML]{FF9341}{44.52}         & \cellcolor[HTML]{FF9442}{44.24} & \cellcolor[HTML]{FF923F}{44.92}        & \cellcolor[HTML]{FFA057}{39.35}        & \cellcolor[HTML]{FFA560}{37.29}         & \cellcolor[HTML]{FFA25A}{38.51} \\
        \lm-13b   &                            & \cellcolor[HTML]{FFEADA}{8.46} & \cellcolor[HTML]{FF7F1C}{53.18} & \cellcolor[HTML]{FF7105}{\uline{58.68}} & \cellcolor[HTML]{FF8D35}{47.30}         & \cellcolor[HTML]{FF8D35}{47.39} & \cellcolor[HTML]{FF903A}{46.06}        & \cellcolor[HTML]{FF9B4F}{41.25}        & \cellcolor[HTML]{FF9C51}{40.76}         & \cellcolor[HTML]{FF9B4F}{41.20} \\
        \cl-7b    &                            & \cellcolor[HTML]{FFFEFE}{0.03} & \cellcolor[HTML]{FF760D}{56.59} & \cellcolor[HTML]{FF6F00}{\uline{59.87}} & \cellcolor[HTML]{FF8B32}{47.91}         & \cellcolor[HTML]{FF8B32}{48.01} & \cellcolor[HTML]{FF8C33}{47.72}        & \cellcolor[HTML]{FF9543}{44.06}        & \cellcolor[HTML]{FF994A}{42.40}         & \cellcolor[HTML]{FF9C50}{41.05} \\
        \cl-13b   & \multirow{-7}{*}{GLEU}     & \cellcolor[HTML]{FFFDFC}{0.60} & \cellcolor[HTML]{FF801E}{52.62} & \cellcolor[HTML]{FF7A15}{\uline{54.90}} & \cellcolor[HTML]{FF9442}{44.32}         & \cellcolor[HTML]{FF9747}{43.04} & \cellcolor[HTML]{FF913D}{45.48}        & \cellcolor[HTML]{FF9E54}{39.97}        & \cellcolor[HTML]{FFA35D}{37.93}         & \cellcolor[HTML]{FFA865}{36.02} \\ \midrule
        \ic-1b    &                            & \cellcolor[HTML]{FFFDFC}{0.60} & \cellcolor[HTML]{FFFAF7}{1.80}  & \cellcolor[HTML]{FFF7F1}{3.10}          & \cellcolor[HTML]{FFF5EE}{3.80}          & \cellcolor[HTML]{FFF6EF}{3.65}  & \cellcolor[HTML]{FFF5EE}{3.95}         & \cellcolor[HTML]{FFF4EC}{\uline{4.35}} & \cellcolor[HTML]{FFF4EC}{4.30}          & \cellcolor[HTML]{FFF8F2}{2.90}  \\
        \cg-2b-nl &                            & \cellcolor[HTML]{FFFFFF}{0.00} & \cellcolor[HTML]{FFFFFF}{0.00}  & \cellcolor[HTML]{FFFEFD}{0.25}          & \cellcolor[HTML]{FFFEFE}{0.20}          & \cellcolor[HTML]{FFFEFE}{0.15}  & \cellcolor[HTML]{FFFEFD}{0.30}         & \cellcolor[HTML]{FFFEFD}{\uline{0.40}} & \cellcolor[HTML]{FFFEFD}{0.35}          & \cellcolor[HTML]{FFFEFD}{0.25}  \\
        \cg-6b-nl &                            & \cellcolor[HTML]{FFFFFF}{0.00} & \cellcolor[HTML]{FFFEFE}{0.10}  & \cellcolor[HTML]{FFFEFD}{0.40}          & \cellcolor[HTML]{FFFBF8}{1.55}          & \cellcolor[HTML]{FFFBF7}{1.65}  & \cellcolor[HTML]{FFFAF7}{\uline{1.70}} & \cellcolor[HTML]{FFFBF9}{1.35}         & \cellcolor[HTML]{FFFAF7}{1.70}          & \cellcolor[HTML]{FFFBF9}{1.25}  \\
        \lm-7b    &                            & \cellcolor[HTML]{FFFDFD}{0.45} & \cellcolor[HTML]{FFDCC2}{14.25} & \cellcolor[HTML]{FFD4B2}{\uline{17.85}} & \cellcolor[HTML]{FFDEC4}{13.65}         & \cellcolor[HTML]{FFDFC6}{13.20} & \cellcolor[HTML]{FFE2CB}{12.05}        & \cellcolor[HTML]{FFEBDC}{8.15}         & \cellcolor[HTML]{FFEDDF}{7.40}          & \cellcolor[HTML]{FFEBDC}{8.00}  \\
        \lm-13b   &                            & \cellcolor[HTML]{FFF7F2}{3.05} & \cellcolor[HTML]{FFD8BA}{16.20} & \cellcolor[HTML]{FFCDA7}{\uline{20.50}} & \cellcolor[HTML]{FFD5B5}{17.30}         & \cellcolor[HTML]{FFD6B7}{16.90} & \cellcolor[HTML]{FFD8BB}{15.85}        & \cellcolor[HTML]{FFE1CB}{12.10}        & \cellcolor[HTML]{FFE4CF}{11.10}         & \cellcolor[HTML]{FFE5D1}{10.60} \\
        \cl-7b    &                            & \cellcolor[HTML]{FFFEFE}{0.05} & \cellcolor[HTML]{FFCBA3}{21.55} & \cellcolor[HTML]{FFC69A}{\uline{23.65}} & \cellcolor[HTML]{FFD1AE}{18.95}         & \cellcolor[HTML]{FFCEA9}{20.00} & \cellcolor[HTML]{FFD0AC}{19.30}        & \cellcolor[HTML]{FFDABE}{15.15}        & \cellcolor[HTML]{FFDCC1}{14.35}         & \cellcolor[HTML]{FFDFC6}{13.30} \\
        \cl-13b   & \multirow{-7}{*}{ACC}      & \cellcolor[HTML]{FFFDFC}{0.50} & \cellcolor[HTML]{FFD1AE}{18.85} & \cellcolor[HTML]{FFCFAB}{\uline{19.65}} & \cellcolor[HTML]{FFD8BB}{15.95}         & \cellcolor[HTML]{FFD8BB}{15.95} & \cellcolor[HTML]{FFD6B6}{16.95}        & \cellcolor[HTML]{FFE6D3}{10.25}        & \cellcolor[HTML]{FFE7D5}{9.70}          & \cellcolor[HTML]{FFE8D7}{9.20}  \\ \bottomrule
    \end{tabular}
    }
\end{table}

Table \ref{table_icl_crg}, \ref{table_icl_cmg} and Table \ref{table_icl_jitcu} show the performance of different LLMs with different settings on the selected code-change-related tasks.
Due to space limitations, for CMG, we only present the experiment results on the Python and Java datasets.
This is because
Python and Java are the most popular programming languages to date.
We present all the experiment results of LLM-ICLs on CMG in Appendix \ref{sec_appendix}.

\subsubsection{Ability of Directly Applying LLM}

We observe that the performance of the LLMs is poor without examples across LLMs and tasks.
One possible reason is that code changes are different from the pre-training data (e.g., code) of LLMs, indicating that LLMs do not have the knowledge related to code change.
With one example provided, the performance of LLMs generally drastically improves.
This indicates that examples in prompt can generally improve the performance of LLMs and the ability to understand code changes can be stimulated via examples.
However, \ic -1b experienced performance fluctuations after increasing the number of examples.
The reason may be that \ic -1b has relatively fewer parameters, and it is still challenging for it to understand and capture the content and associations of multiple examples in the input~\cite{kaplan2020scaling}.

\rqbox{\textbf{Finding 1}:
    LLMs lack the knowledge specific to code changes. Providing examples in the prompt can generally improve their performance on the tasks related to code changes.}\label{finding1}

\subsubsection{Impact of Different Numbers of Examples}

We observe that as the number of examples in prompt increases, the performance of LLMs increases and then decreases across LLMs and tasks.
For instance, on JITCU, the GLEU/ACC scores of \lm /\cl~increase at first, until the number of examples in the prompt is 2 and then decrease after adding more examples.
On CRG, the BLEU scores of \cl-7b achieve the highest 5.04 when the number of examples in the prompt is 7.
This may result from the limitation of input context length.
As indicated in Section~\ref{sec_design_implementation}, we allocate 1024 tokens for the examples in the prompt.
This indicates that when there are too many examples, we often need to truncate the length of each example in tasks.
For example, after being tokenized by the tokenizer of \cl, the average length of the samples in the training sets has more than 211/184 tokens on the CRG and JITCU tasks, respectively.
With more examples, we need to truncate the examples by removing more tokens from the examples.
This hinders LLMs from understanding the information from each example and can result in a drop in performance.
And, due to the varying length distributions of data across different tasks, the number of examples required for a model to achieve the best performance varies for different tasks.

\rqbox{\textbf{Finding 2}:
    More examples do not always lead to better performance. The effectiveness of LLMs often depends on the distribution of data lengths in the task and the context length allocated to the model.
}\label{finding2}

\subsubsection{Impact of Model Size}

Prior studies showed that within the same LLM family, larger models are associated with better performance \cite{roziere2023code,nijkamp2022codegen,touvron2023llama}.
This also applies to some extent to tasks related to code change.
For example, in the \lm~and \cg~families, better performance is achieved by larger models.
One possible reason is that the larger models can have a broader understanding and capacity to integrate more context effectively due to the vast number of parameters, thus performing better in these families.
\textbf{However, in the \cl~family, better performance on three tasks is achieved by smaller models.}
A possible reason can be that these smaller models might have just the right capacity to capture the essential patterns without being bogged down by excessive complexity~\cite{tay2022scaling,tay2021scale}.
Therefore, the observed performance differences highlight that model size and performance can be task-specific and that bigger is not always better when it comes to using ICL within different LLM families.

\rqbox{\textbf{Finding 3}:
    Within the same LLM family, larger models do not always have better performance.
}\label{finding3}

\subsubsection{Impact of Model Family}

We also find that the \textbf{\cl~family performs the best across code-change-related tasks compared to other model families}.
This may be because \cl~is based on \lm~and has been pre-trained on a large amount of (1) code data and (2) code-related natural language data, enabling it to better understand the information in code and natural language at the same time.
For instance, \textbf{on CRG, CMG-Java, CMG-Python, and JITCU, the best LLM-ICLs are all \cl-7b.}
\cl-7b demonstrates the strongest learning capabilities.
As the number of examples increases, the performance of \cl-7b improves the most on average.
For example, on the CMG-Java, the best \cl-7b with examples outperforms its 0 shot setting by 3.47 times.
In comparison, \lm-13b outperforms its 0 shot setting by 2.48 times.
This suggests that although the model may struggle to comprehend the task without example, \cl-7b can rapidly learn and capture task-relevant features when provided with examples.
These findings highlight the advantage of \cl-7b in adapting to new tasks.
\textbf{When applying LLMs to new code-change-related tasks, we recommend using \cl~family, especially \cl-7b.}

\rqbox{\textbf{Finding 4}:
    The \cl~family, especially \cl-7b, performs the best in the selected tasks related to code changes.
}\label{finding4}

\subsection{RQ2: How Do LLMs Perform When Applying Parameter-Efficient Fine-Tuning Techniques on Code-Change-Related Tasks?}
\label{rq_peft}

\begin{table}[!htbp]
    \caption{{[}RQ2{]} - Performance of LLM-PEFTs on the CRG and JITCU tasks. The darker color of the cells means better performance.}
    \label{table_peft_crg_jitcu}
    \centering
    \resizebox{0.8\linewidth}{!}{
    \begin{tabular}{lcc|cccc}
        \toprule
        \multicolumn{1}{c}{}                                 & \multicolumn{2}{c|}{\textbf{CRG}}    & \multicolumn{4}{c}{\textbf{JITCU}}                                                                                                                                         \\ \cmidrule(l){2-7}
        \multicolumn{1}{c}{}                                 & \multicolumn{2}{c|}{\textbf{BLEU-4}} & \multicolumn{2}{c}{\textbf{GLEU}}  & \multicolumn{2}{c}{\textbf{ACC}}                                                                                                      \\ \cmidrule(l){2-7}
        \multicolumn{1}{c}{\multirow{-3}{*}{\textbf{Model}}} & \textbf{LORA}                        & \textbf{Prefix}                    & \textbf{LORA}                    & \textbf{Prefix}                 & \textbf{LORA}                   & \textbf{Prefix}                \\ \midrule
        \ic-1b                                               & \cellcolor[HTML]{FFC08F}{2.50}       & -                                  & \cellcolor[HTML]{FFAA69}{38.23}  & -                               & \cellcolor[HTML]{FFE4CF}{12.10} & -                              \\
        \cg-2b-nl                                            & \cellcolor[HTML]{FFFEFD}{0.27}       & \cellcolor[HTML]{FFEFE3}{0.61}     & \cellcolor[HTML]{FFF8F3}{0.26}   & \cellcolor[HTML]{FFFCFA}{1.13}  & \cellcolor[HTML]{FFFDFC}{0.55}  & \cellcolor[HTML]{FFFEFE}{0.05} \\
        \cg-6b-nl                                            & \cellcolor[HTML]{FFE3CD}{1.11}       & \cellcolor[HTML]{FFE9D8}{0.87}     & \cellcolor[HTML]{FFF6EF}{4.00}   & \cellcolor[HTML]{FFBE8C}{29.36} & \cellcolor[HTML]{FFFCFA}{1.15}  & \cellcolor[HTML]{FFF7F1}{3.35} \\
        \lm-7b                                               & \cellcolor[HTML]{FF6F00}{5.74}       & \cellcolor[HTML]{FFE2CD}{1.12}     & \cellcolor[HTML]{FF7307}{63.30}  & \cellcolor[HTML]{FFFEFE}{0.25}  & \cellcolor[HTML]{FFB780}{32.25} & \cellcolor[HTML]{FFFFFF}{0.00} \\
        \lm-13b                                              & \cellcolor[HTML]{FF7A13}{5.29}       & \cellcolor[HTML]{FFC293}{2.42}     & \cellcolor[HTML]{FF760D}{61.81}  & \cellcolor[HTML]{FFFFFF}{0.00}  & \cellcolor[HTML]{FFB780}{32.45} & \cellcolor[HTML]{FFFFFF}{0.00} \\
        \cl-7b                                               & \cellcolor[HTML]{FF994B}{4.04}       & \cellcolor[HTML]{FFF6F0}{0.33}     & \cellcolor[HTML]{FF6F00}{65.23}  & \cellcolor[HTML]{FFFEFE}{0.06}  & \cellcolor[HTML]{FFB378}{34.40} & \cellcolor[HTML]{FFFFFF}{0.00} \\
        \cl-13b                                              & \cellcolor[HTML]{FF7205}{5.61}       & \cellcolor[HTML]{FFE1CB}{1.17}     & \cellcolor[HTML]{FF7103}{64.22}  & \cellcolor[HTML]{FFFEFD}{0.27}  & \cellcolor[HTML]{FFB176}{34.90} & \cellcolor[HTML]{FFFFFF}{0.00} \\ \bottomrule
    \end{tabular}
    }
\end{table}

\begin{table}[!htbp]
    \caption{{[}RQ2{]} - Performance of LLM-PEFTs on the CMG task. The darker color of the cells means better performance.}
    \label{table_peft_cmg}
    \centering
    \resizebox{\textwidth}{!}{
    \begin{tabular}{lcccccccccc}
        \toprule
        \multicolumn{1}{c}{}                                  & \multicolumn{2}{c}{\textbf{Java}} & \multicolumn{2}{c}{\textbf{C\#}} & \multicolumn{2}{c}{\textbf{CPP}} & \multicolumn{2}{c}{\textbf{Python}} & \multicolumn{2}{c}{\textbf{JavaScript}}                                                                                                                                                                        \\ \cmidrule(l){2-11}
        \multicolumn{1}{c}{\multirow{-2}{*}{\textbf{Models}}} & \textbf{LORA}                     & \textbf{Prefix}                  & \textbf{LORA}                    & \textbf{Prefix}                     & \textbf{LORA}                           & \textbf{Prefix}                & \textbf{LORA}                   & \textbf{Prefix}                & \textbf{LORA}                   & \textbf{Prefix}                \\ \midrule
        \ic-1b                                                & \cellcolor[HTML]{FFC9A0}{5.13}    & -                                & \cellcolor[HTML]{FFCBA3}{4.95}   & -                                   & \cellcolor[HTML]{FFC599}{5.54}          & -                              & \cellcolor[HTML]{FFC598}{5.57}  & -                              & \cellcolor[HTML]{FFC69A}{5.47}  & -                              \\
        \cg-2b-nl                                             & \cellcolor[HTML]{FFE7D6}{2.23}    & \cellcolor[HTML]{FFEBDC}{1.87}   & \cellcolor[HTML]{FFDDC2}{3.27}   & \cellcolor[HTML]{FFE2CD}{2.71}      & \cellcolor[HTML]{FFDCC1}{3.37}          & \cellcolor[HTML]{FFD9BC}{3.60} & \cellcolor[HTML]{FFD7B9}{3.77}  & \cellcolor[HTML]{FFDABE}{3.52} & \cellcolor[HTML]{FFDABD}{3.55}  & \cellcolor[HTML]{FFD6B7}{3.89} \\
        \cg-6b-nl                                             & \cellcolor[HTML]{FFD9BB}{3.65}    & \cellcolor[HTML]{FFE2CB}{2.78}   & \cellcolor[HTML]{FFCDA7}{4.77}   & \cellcolor[HTML]{FFE3CE}{2.62}      & \cellcolor[HTML]{FFCBA4}{4.94}          & \cellcolor[HTML]{FFDFC6}{3.08} & \cellcolor[HTML]{FFC89E}{5.27}  & \cellcolor[HTML]{FFCDA6}{4.81} & \cellcolor[HTML]{FFC9A0}{5.14}  & \cellcolor[HTML]{FFD9BC}{3.59} \\
        \lm-7b                                                & \cellcolor[HTML]{FF7F1C}{12.30}   & \cellcolor[HTML]{FFECDE}{1.77}   & \cellcolor[HTML]{FF8222}{11.97}  & \cellcolor[HTML]{FFF4EB}{1.05}      & \cellcolor[HTML]{FF8C34}{11.02}         & \cellcolor[HTML]{FFFDFC}{0.13} & \cellcolor[HTML]{FF7308}{13.35} & \cellcolor[HTML]{FFFEFE}{0.05} & \cellcolor[HTML]{FF6F00}{13.84} & \cellcolor[HTML]{FFFEFE}{0.03} \\
        \lm-13b                                               & \cellcolor[HTML]{FF8527}{11.73}   & \cellcolor[HTML]{FFF7F1}{0.72}   & \cellcolor[HTML]{FF862A}{11.57}  & \cellcolor[HTML]{FFF4EC}{1.00}      & \cellcolor[HTML]{FF9341}{10.31}         & \cellcolor[HTML]{FFF7F2}{0.69} & \cellcolor[HTML]{FF7912}{12.85} & \cellcolor[HTML]{FFF6EF}{0.84} & \cellcolor[HTML]{FF7308}{13.39} & \cellcolor[HTML]{FFF4EC}{1.03} \\
        \cl-7b                                                & \cellcolor[HTML]{FF770E}{13.06}   & \cellcolor[HTML]{FFF3EA}{1.12}   & \cellcolor[HTML]{FF8121}{12.05}  & \cellcolor[HTML]{FFF2E8}{1.20}      & \cellcolor[HTML]{FF8C34}{11.01}         & \cellcolor[HTML]{FFF5EE}{0.89} & \cellcolor[HTML]{FF740A}{13.37} & \cellcolor[HTML]{FFF8F3}{0.61} & \cellcolor[HTML]{FF6F01}{13.79} & \cellcolor[HTML]{FFF6EF}{0.85} \\
        \cl-13b                                               & \cellcolor[HTML]{FF8323}{11.93}   & \cellcolor[HTML]{FFEDE0}{1.68}   & \cellcolor[HTML]{FF8120}{12.12}  & \cellcolor[HTML]{FFE0C8}{2.98}      & \cellcolor[HTML]{FF8426}{11.77}         & \cellcolor[HTML]{FFE4D0}{2.51} & \cellcolor[HTML]{FF7811}{12.92} & \cellcolor[HTML]{FFD6B6}{3.93} & \cellcolor[HTML]{FF6F01}13.87   & \cellcolor[HTML]{FFEEE1}{1.60} \\ \bottomrule
    \end{tabular}
     }
\end{table}

Table \ref{table_peft_crg_jitcu} and Table \ref{table_peft_cmg} show the results of LLM-PEFTs on the three code-change-related tasks.

\subsubsection{Comparison Between \lora~and Prefix-Tuning}

We observe that when performing PEFT on code change-related tasks, LLMs tuned with \lora~results in significantly better performance compared to LLMs tuned with prefix-tuning.
For example, on CRG and CMG, the average BLEU scores of \cl -13b using \lora~are 5.61 and 12.52, respectively; while those of \cl -13b using prefix-tuning are 1.17, and 2.54, respectively.
This indicates that \lora~can help LLMs learn more knowledge related to code changes compared to prefix-tuning.
These results are also in line with the different mechanisms of \lora~and prefix-tuning.
Specifically, \lora~updates the self-attention modules in LLMs, making it relatively easy to learn new knowledge that is not well covered by the pre-training data. However prefix-tuning, which only prepends some trainable vectors to the input of each layer, plays a similar role to soft prompts, i.e., stimulating the learned knowledge in LLMs.
Considering existing LLMs are not specially trained for code-change-related tasks and have limited knowledge of code changes, it is reasonable that \lora~outperforms prefix-tuning.

\begin{table}[!htbp]
  \centering
  \caption{{[}RQ2{]} -- An example of the CMG task.}
  \label{table_peft_example}
  \scriptsize
  \begin{tabular}{l|l}
    \hline
    \textbf{Diff}                        &
    \begin{tabular}[c]{@{}l@{}}
      \begin{lstlisting}[language=Java]
public class AnalystWorker implements Runnable { 
   /** Open a single point channel to the broker
    *    to receive high-priority requests immediately 
    */
   private synchronized void openSideChannel () { 
+   if (sideChannelOpen)
+       return; 
+    
    LOG.info("Opening side channel for single point requests.");
    new Thread(() ->{
        sideChannelOpen = true;
  \end{lstlisting}
    \end{tabular}                                           \\ \hline
    \textbf{Gold}                        & resolve race condition where two side channels could open.         \\ \hline
    \textbf{Llama 2-13b + LoRA}          & Fixed a bug in AnalystWorker where it would re-open a side channel \\
    \textbf{Llama 2-13b + Prefix-Tuning} & data class in it. java, to.                                        \\ \hline
  \end{tabular}
\end{table}

Table \ref{table_peft_example} presents an example of CMG and the commit messages generated by two \lm-13b models that are fine-tuned with \lora~and prefix-tuning, respectively.
We observe that the commit message generated by \lm-13b with \lora~successfully captures the changed parts in the diff, while the \lm-13b with prefix-tuning only generates some low-quality keywords.
Similar phenomena are also reported by prior studies \cite{ding2023parameter,chen2022inducer,chen2022revisiting}.

\rqbox{\textbf{Finding 5}:
    When applying LLM-PEFTs, tuning LLMs using \lora~achieve a significantly better performance compared to those using prefix-tuning.
}\label{finding5}

\subsubsection{Impact of Model Size}

When tuning LLMs with \lora, the performance difference between large and small models in the same LLM family is not significant.
For example, on CRG, \cl-13b outperforms \cl-7b.
While, on JITCU, smaller models in the \lm~family and the \cl~family outperform the larger models in terms of GLEU scores.
One possible reason is that smaller models may have just sufficient adaptability and efficiency when fine-tuning with \lora, allowing them to capture relevant features effectively for specific tasks.
The complexity and capacity of larger models may not provide additional benefits on these tasks related to code changes and might even introduce unnecessary complexity.
This means that when tuning LLM using \lora~on tasks related to code changes, developers should consider both the bigger and smaller models within the same model family at the same time.

\rqbox{\textbf{Finding 6}:
    When tuning LLMs using \lora, larger models do not necessarily have better performance, even within the same LLM family.
}\label{finding6}

\subsubsection{Impact of Model Family}

We observe that when tuning LLMs with \lora, \lm~and \cl~families are the best-performing LLMs on the three code-change-related tasks.
Specifically, on CRG and CMG-Python, the \lm~performs the best;
however, on other tasks, \cl~performs better.
Nonetheless, the performance gap between the two model families is not significant.
For example, on CRG, the best-performing \lm~model (\lm-7b) has a BLEU score of 5.74, which is only approximately 2.32\% higher than the best-performing \cl~model (\cl-13b) with a score of 5.61.
One possible reason is that, though \cl~is based on \lm~and has been pre-trained on a large-scale code corpus, tuning LLMs using \lora~on code-change-related tasks requires the model to learn new knowledge that is not well covered by the pre-training data.
This makes these two model families perform similarly on code-change-related tasks.
\textbf{We recommend exploring both Llama and \cl~families when tuning LLMs using \lora}.

\rqbox{\textbf{Finding 7}:
    When tuning LLM using \lora~on tasks related to code changes, utilizing models specifically pre-trained on code-related tasks (such as \cl) offers limited benefits.
    \lm~and \cl~families are the best-performing LLMs.
}\label{finding7}

\subsection{RQ3: How Do LLMs Perform on Code Change-Related Tasks Compared to Small Pre-trained Models?}\label{rq_compare}

\begin{table*}[!htbp]
    \caption{{[}RQ3{]} - Performance of the small pre-trained models after fully fine-tuned on code-change-related tasks. The deeper the color, the better the performance.}
    \label{table_small}
    \centering
    \resizebox{0.8\linewidth}{!}{
    \begin{tabular}{lcccccccc}
        \toprule
        \multicolumn{1}{c}{}                                 & \textbf{CRG}                 & \multicolumn{5}{c}{\textbf{CMG}} & \multicolumn{2}{c}{\textbf{JITCU}}                                                                                                                                                                 \\ \cmidrule(l){2-9}
        \multicolumn{1}{c}{}                                 &                              & \textbf{CPP}                     & \textbf{C\#}                       & \textbf{Java}                 & \textbf{JavaScript}           & \textbf{Python}               &                               &                               \\ \cmidrule(lr){3-7}
        \multicolumn{1}{c}{\multirow{-3}{*}{\textbf{Model}}} & \multirow{-2}{*}{BLEU}       & BLEU                             & BLEU                               & BLEU                          & BLEU                          & BLEU                          & \multirow{-2}{*}{GLEU}        & \multirow{-2}{*}{ACC}         \\ \midrule
        Codet5                             & \cellcolor[HTML]{FEF8F4}0.38 & \cellcolor[HTML]{FEF8F4}2.85     & \cellcolor[HTML]{FEF8F4}3.73       & \cellcolor[HTML]{FCE4D6}5.33  & \cellcolor[HTML]{FCE4D6}7.32  & \cellcolor[HTML]{FCE4D6}5.32  & \cellcolor[HTML]{ED7F3B}59.14 & \cellcolor[HTML]{F2A068}17.90 \\
        \cc                                & \cellcolor[HTML]{ED7F3B}5.30 & \cellcolor[HTML]{F8CBAD}12.99    & \cellcolor[HTML]{ED7F3B}19.53      & \cellcolor[HTML]{F2A372}15.90 & \cellcolor[HTML]{ED7F3B}17.61 & \cellcolor[HTML]{F2A372}14.57 & \cellcolor[HTML]{ED7F3B}68.32 & \cellcolor[HTML]{ED7F3B}29.50 \\ \bottomrule
    \end{tabular}
    }
\end{table*}

\begin{figure}[htbp]
\centering
\begin{minipage}[c]{0.42\textwidth}
\centering
\includegraphics[width=\textwidth]{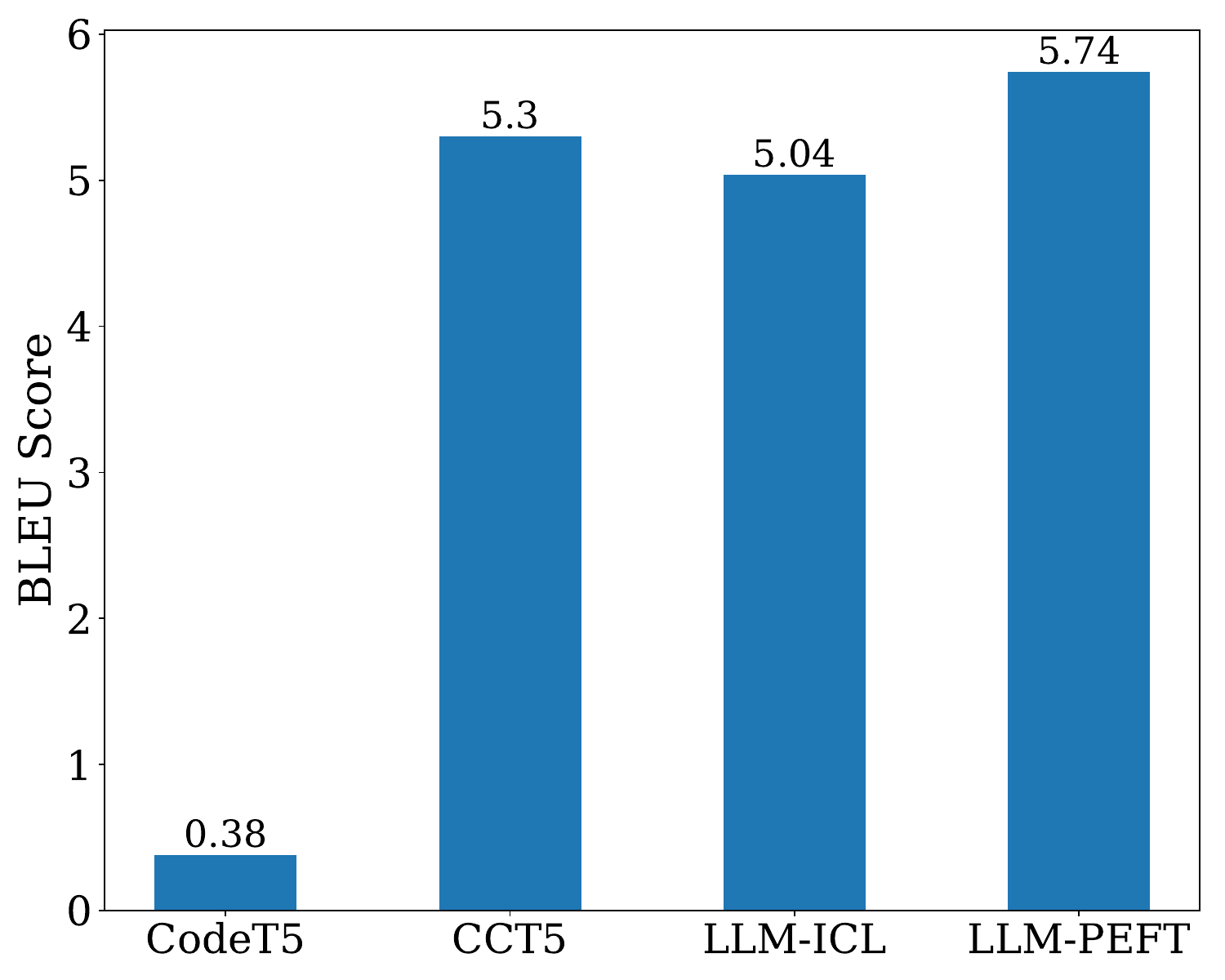}
\subcaption{BLEU on CRG}\label{figure_comparison_crg}
\label{fig_compare_crg}
\end{minipage}
\begin{minipage}[c]{0.42\textwidth}
\centering
\includegraphics[width=\textwidth]{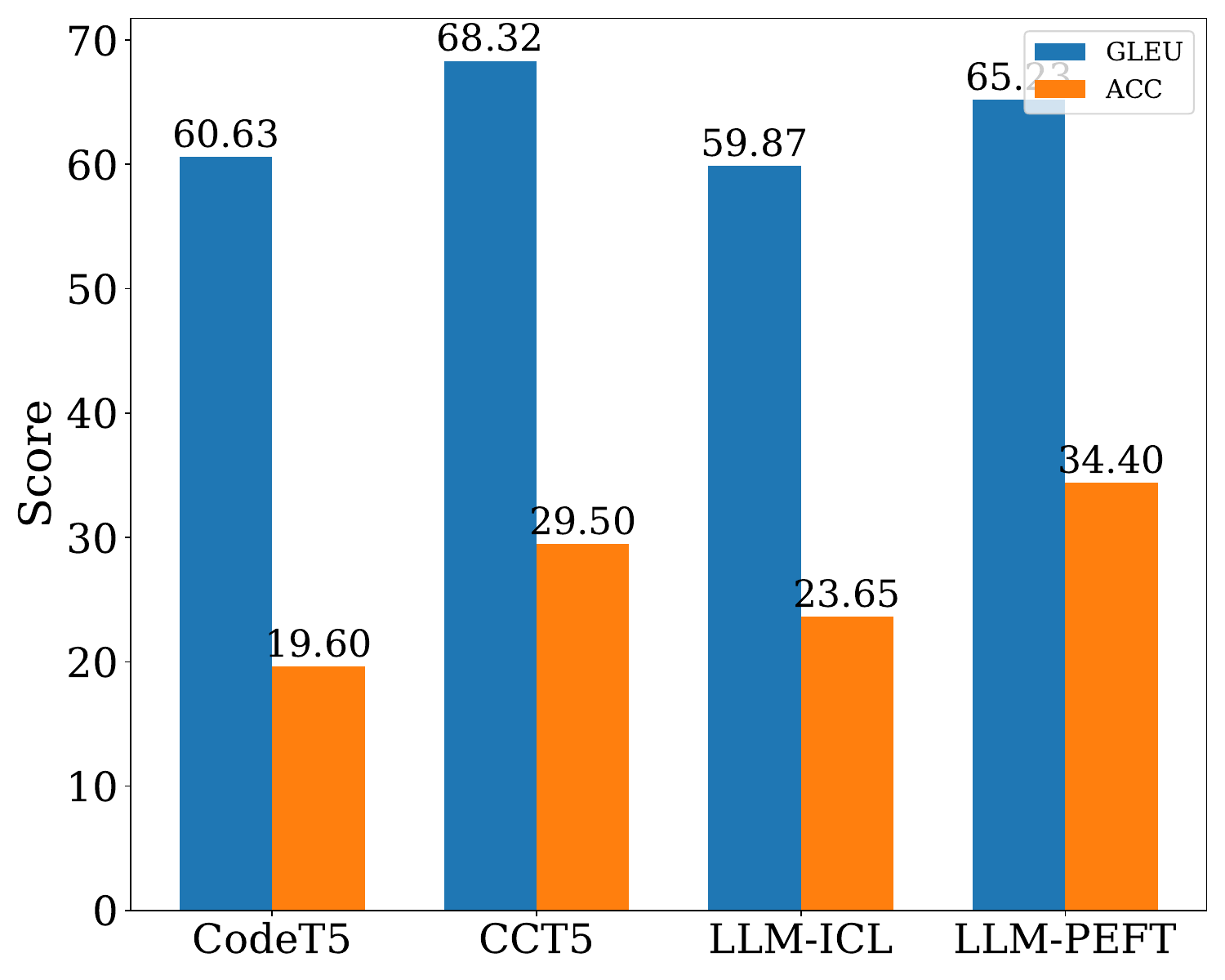}
\subcaption{GLEU and ACC on JITCU}\label{figure_comparison_jitcu}
\label{fig_compare_jit}
\end{minipage}
\caption{{[}RQ3{]} - Comparison between the best performing LLM-ICL, LLM-PEFT, and Small Pre-trained Models on CRG and JITCU}
\end{figure}

\begin{figure}[htbp]
    \centering
    \includegraphics[width=0.85\textwidth]{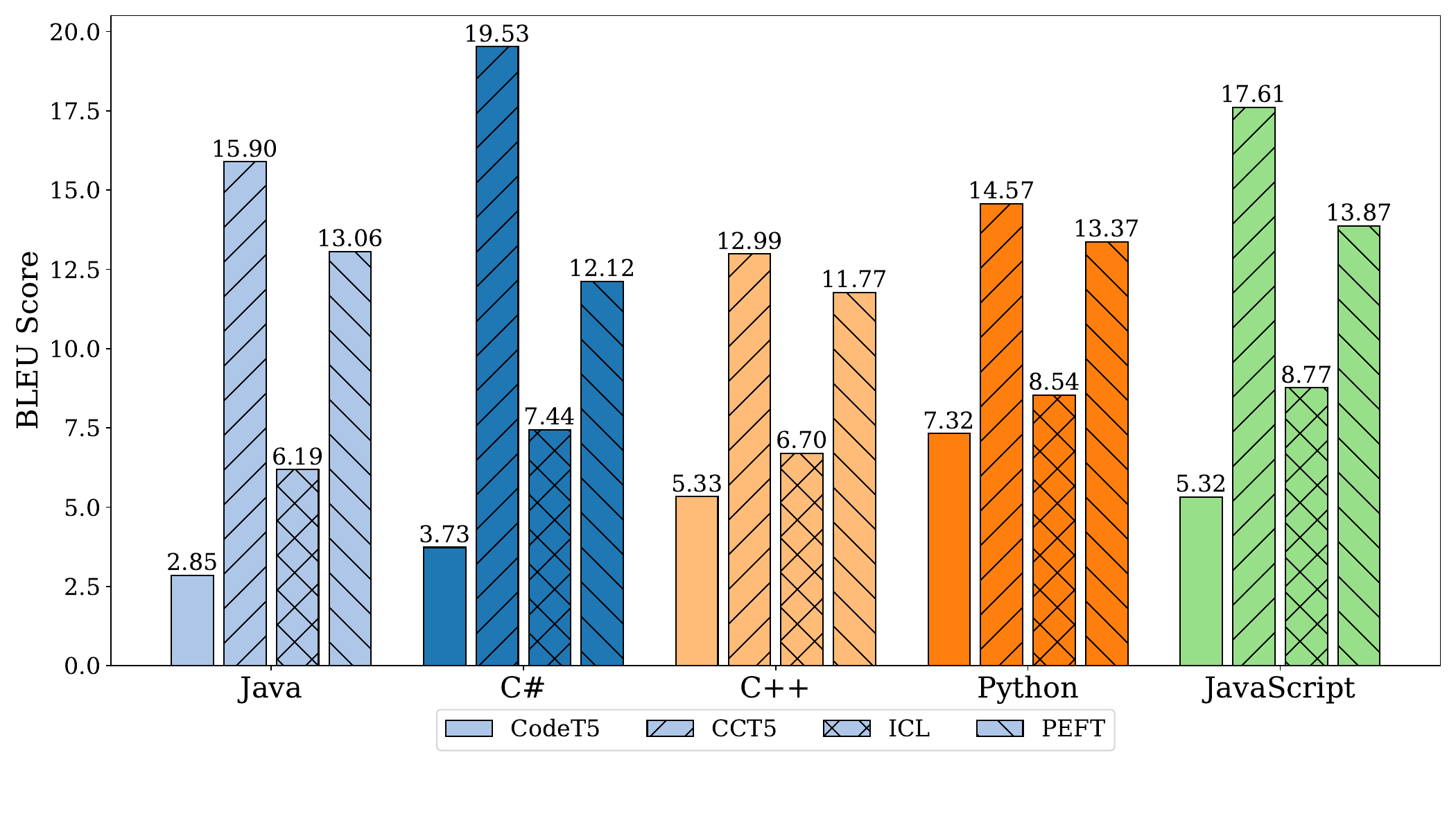}
    \caption{{[}RQ3{]} - Comparison between the best performing LLM-ICL, LLM-PEFT, and Small Pre-trained Models on CMG}
    \label{figure_comparison_cmg}
\end{figure}

Table \ref{table_small} presents the performance of the \sm s that are fully fine-tuned on the three code-change-related tasks.
We observe that \cc~outperforms CodeT5 on all tasks.
This is because \cc~is pre-trained with code changes and code-change-related natural language descriptions, which makes it good at handling code-change-related tasks.
In contrast, CodeT5 is only pre-trained with code and code-related descriptions.
To better show the differences between LLMs and \sm s, we compare the best-performing LLM-ICL, LLM-PEFT, and \sm s on CRG, CMG, and JITCU.
Figure \ref{figure_comparison_crg}, Figure \ref{figure_comparison_cmg}, and Figure \ref{figure_comparison_jitcu} visualize the performances of the best performing LLM-ICL, LLM-PEFT, and \sm s on CRG, CMG, and JITCU, respectively.

\subsubsection{Comparison between LLM-ICLs and Small Pre-trained Models}

The best-performing LLM-ICLs are similar to or better than CodeT5 on the tasks related to code changes.
Specifically, the best-performing LLM-ICLs can statistically significantly outperform CodeT5 on CRG, the C++, C\#, and Python sub-datasets on CMG.
For example, \cl-7b with 7 examples outperforms CodeT5 by 4.66 on CRG.
Besides, there is no significant difference between the best-performing LLM-ICL and CodeT5 on JITCU in terms of GLEU.
For example, CodeT5 only outperforms the best-performing LLM-ICL (i.e., \cl-7b with 2 examples) by 0.76 on JITCU.
These results indicate that LLMs are promising on code-change-related tasks.

The best-performing LLM-ICLs are statistically significantly inferior to \cc~on all tasks.
Specifically, \cc~outperforms \cl-7b with 7 examples on CRG by 0.26.
Recall that \cc~are pre-trained with code-change-oriented objectives and fine-tuned with task-specific datasets, and the parameters have learned the knowledge related to code changes through updates.
This further motivates us to compare \peft~with \sm~on the tasks related to code changes, both of which can update model parameters.

\rqbox{\textbf{Finding 10}:
    The best-performing LLM-ICLs are similar or better than small models pre-trained with code, but inferior to small models pre-trained with code changes on the tasks related to code changes.
}\label{finding10}

\subsubsection{Comparison between LLM-ICLs and LLM-PEFTs}

The best-performing LLM-PEFTs (LLMs tuned with \lora) outperform the best-performing LLM-ICLs across all tasks, with particularly significant differences on CMG and JITCU.
Specifically, the best-performing LLM-PEFT performs slightly better than the best-performing LLM-ICL on CRG;
the best-performing LLM-PEFTs outperform the best-performing LLM-ICLs by 123.55\%, 89.67\%, 77.53\%, 60.96\% and 103.67\% on Java, C\#, C++, Python and JavaScript sub-datasets of CMG respectively;
the best-performing LLM-PEFTs outperform the best-performing LLM-ICLs by 8.95\% and 47.45\% in terms of GLEU and ACC on JITCU.
This means that if model parameters are allowed to be adjusted, the performance of LLMs on code change-related tasks can be significantly improved.

\rqbox{\textbf{Finding 11}:
    Tuning LLM with \lora~can significantly improve the performance of LLMs on code-change-related tasks.
}\label{finding11}

\subsubsection{Comparison between LLM-PEFTs and Small Pre-trained Models}

\begin{table*}[!htbp]
    \caption{{[}RQ2{]} - The number of parameters updated when fine-tuning each LLM with PEFT}\label{table_peft_parameters}
    \centering
    \resizebox{\textwidth}{!}{
        \begin{tabular}{l|ccccccc}
            \hline
            Method        & \ic-1b    & \cg-2b-nl & \cg-6b-nl & \lm-7b     & \lm-13b   & \cl-7b & \cl-13b \\ \hline
            \lora          & 3,15M(0.23\%) & 2.62M(0.09\%) & 4.3M(0.06\%)  & 8,38M(0.12\%)  & 13,1M(0.10\%) & 8.39M(0.12\%) & 13.1M(0.10\%)  \\
            prefix-tuning & -             & 1.64M(0.06\%) & 2.7M(0.04\%)  & 2,62M( 0.04\%) & 4.1M(0.03\%)  & 2.62M(0.04\%) & 4.1M(0.03\%)   \\ \hline
        \end{tabular}
    }
\end{table*}

We observe that the best-performing LLM-PEFTs (LLMs tuned with \lora) statistically significantly outperform CodeT5 on all tasks.
Moreover, on CRG, the best-performing LLM-PEFT, i.e., \lm-7B tuned using \lora~statistically significantly outperforms the best fully fine-tuned \sm, i.e., \cc, by 8.3\%.
On JITCU, the best-performing LLM-PEFT, i.e., \cl-13B tuned using \lora, outperforms the fully fine-tuned \cc~by 18.47\% in terms of ACC.
On CMG, there is no significant difference between the best-performing LLM-PEFT and the best fully fine-tuned \sm~(\cc) on C++, Java, and Python sub-datasets, and on the Javascript and C\# sub-datasets, the best-performing LLM-PEFT are statistically significantly inferior to \cc.
These indicate that the best-performing LLM-PEFTs have comparable performance to \cc.

Table \ref{table_peft_parameters} shows the number of parameters that are updated with different PEFT methods.
The number of parameters updated when fine-tuning each LLM with PEFT is fewer than the total number of parameters (i.e., 223M) of the small pre-trained models (as is shown in Table \ref{table_selected_llms}).
Considering that LLMs are not pre-trained with code-change-oriented objectives, we believe the performance of LLMs can be further improved by pre-training on the objectives and data related to code changes.

\rqbox{\textbf{Finding 12}:
    The best-performing LLM-PEFTs can have comparable performance to the best-performing fully fine-tuned \sm s on the tasks related to code changes with fewer changed parameters.
}\label{finding12}

\subsection{RQ4: How Do LLMs Perform With Different Input Formats on Code-Change-Related Tasks?}\label{rq_input}

\begin{table*}[!htbp]
\centering
    \caption{ {[}RQ3{]} - BLEU scores of LLM-ICLs with the code input on CRG. Blue indicates worse performance while red indicates better performance, compared to using diff as input. The deeper the color, the greater the difference.}
    \label{table_code_icl_crg}
    \resizebox{0.9\linewidth}{!}{
    \begin{tabular}{lccccccccc}
        \toprule
        Models        & 0shot                          & 1shot                                  & 2shot                          & 3shot                          & 4shot                                  & 5shot                          & 6shot                          & 7shot                                  & 8shot                                  \\ \midrule
        \ic-1b    & \cellcolor[HTML]{FABCB5}{2.97} & \cellcolor[HTML]{FADFDB}{\uline{3.95}} & \cellcolor[HTML]{FAF0EF}{3.82} & \cellcolor[HTML]{FAEEED}{3.84} & \cellcolor[HTML]{FAF0EE}{3.53}         & \cellcolor[HTML]{FAEDEB}{3.42} & \cellcolor[HTML]{FAECEA}{3.3}  & \cellcolor[HTML]{FAEBE9}{3.23}         & \cellcolor[HTML]{FAF4F3}{3.04}         \\
        \cg-2b-nl & \cellcolor[HTML]{FAF8F7}{0.31} & \cellcolor[HTML]{FAF2F1}{0.5}          & \cellcolor[HTML]{FAF2F1}{0.5}8 & \cellcolor[HTML]{89BDFF}{0.7}  & \cellcolor[HTML]{8FC0FF}{0.88}         & \cellcolor[HTML]{A1CAFF}{1.03} & \cellcolor[HTML]{D4E7FF}{1.32} & \cellcolor[HTML]{FAFAF9}{1.69}         & \cellcolor[HTML]{FAF0EF}{\uline{1.92}} \\
        \cg-6b-nl & \cellcolor[HTML]{FAF1F0}{0.87} & \cellcolor[HTML]{FAF7F6}{0.93}         & \cellcolor[HTML]{FAF7F6}{1.11} & \cellcolor[HTML]{FAF5F4}{1.4}  & \cellcolor[HTML]{FAEDEB}{1.75}         & \cellcolor[HTML]{FADFDB}{2.07} & \cellcolor[HTML]{FAD4D0}{2.57} & \cellcolor[HTML]{FACBC5}{2.78}         & \cellcolor[HTML]{FAD5D1}{\uline{2.95}} \\
        \lm-7b    & \cellcolor[HTML]{FAF7F6}{0.93} & \cellcolor[HTML]{EFF6FF}{4.33}         & \cellcolor[HTML]{FAFEFE}{4.43} & \cellcolor[HTML]{F8FBFF}{4.52} & \cellcolor[HTML]{FAFCFC}{4.54}         & \cellcolor[HTML]{F3F8FF}{4.67} & \cellcolor[HTML]{F3F8FF}{4.67} & \cellcolor[HTML]{D6E8FF}{4.73}         & \cellcolor[HTML]{EDF5FF}{\uline{4.8}}  \\
        \lm-13b   & \cellcolor[HTML]{FAE3E1}{2.34} & \cellcolor[HTML]{FAFDFD}{4.28}         & \cellcolor[HTML]{FAFDFD}{4.45} & \cellcolor[HTML]{FAFBFB}{4.69} & \cellcolor[HTML]{FAFDFD}{\uline{4.71}} & \cellcolor[HTML]{E6F1FF}{4.61} & \cellcolor[HTML]{FAFDFD}{4.71} & \cellcolor[HTML]{FAFDFD}{4.71}         & \cellcolor[HTML]{FAFDFD}{4.71}         \\
        \cl-7b  & \cellcolor[HTML]{FFFFFF}{0.9}  & \cellcolor[HTML]{FAFCFF}{4.53}         & \cellcolor[HTML]{C9E1FF}{4.5}  & \cellcolor[HTML]{FAFBFB}{4.69} & \cellcolor[HTML]{C9E1FF}{4.78}         & \cellcolor[HTML]{D6E8FF}{4.79} & \cellcolor[HTML]{EDF5FF}{4.8}6 & \cellcolor[HTML]{C2DDFF}{4.77}         & \cellcolor[HTML]{F8FBFF}{\uline{4.9}}  \\
        \cl-13b & \cellcolor[HTML]{FAFCFC}{2.32} & \cellcolor[HTML]{C9E1FF}{4.5}7         & \cellcolor[HTML]{C9E1FF}{4.5}7 & \cellcolor[HTML]{FAFDFD}{4.71} & \cellcolor[HTML]{D6E8FF}{4.79}         & \cellcolor[HTML]{EDF5FF}{4.8}3 & \cellcolor[HTML]{D6E8FF}{4.79} & \cellcolor[HTML]{FAFBFB}{\uline{5.02}} & \cellcolor[HTML]{F8FBFF}{4.9}          \\ \bottomrule
    \end{tabular}
    }
\end{table*}

\begin{table}[!htbp]
\centering
    \caption{ {[}RQ3{]} - BLEU scores of LLM-ICLs with the code input format on Python and Java sub-datasets of CMG. Blue indicates worse performance while red indicates better performance, compared to using diff as input. The deeper the color, the greater the difference.}
    \label{table_code_icl_cmg}
    \resizebox{\linewidth}{!}{
    \begin{tabular}{clccccccccc}
        \toprule
        \multicolumn{1}{l}{Lang} & Model     & 0shot                                  & 1shot                          & 2shot                          & 3shot                                  & 4shot                          & 5shot                                  & 6shot                                  & 7shot                                  & 8shot                                  \\ \midrule
                                 & \ic-1b    & \cellcolor[HTML]{E1EEFF}{\uline{3.08}} & \cellcolor[HTML]{FAFCFC}{3}    & \cellcolor[HTML]{F4F9FF}{2.62} & \cellcolor[HTML]{F5F9FF}{2.74}         & \cellcolor[HTML]{FAF7F7}{2.66} & \cellcolor[HTML]{F8FBFF}{2.75}         & \cellcolor[HTML]{FFFFFF}{2.79}         & \cellcolor[HTML]{F1F7FF}{2.76}         & \cellcolor[HTML]{FAF7F7}{2.66}         \\
                                 & \cg-2b-nl & \cellcolor[HTML]{FDFEFF}{1.88}         & \cellcolor[HTML]{F1F7FF}{1.98} & \cellcolor[HTML]{E7F1FF}{2}    & \cellcolor[HTML]{E7F1FF}{2}.11         & \cellcolor[HTML]{E7F1FF}{2}.28 & \cellcolor[HTML]{E7F1FF}{2}.29         & \cellcolor[HTML]{E7F1FF}{2}.29         & \cellcolor[HTML]{E7F1FF}{\uline{2.36}} & \cellcolor[HTML]{E7F1FF}{2}.33         \\
                                 & \cg-6b-nl & \cellcolor[HTML]{E7F1FF}{2}.46         & \cellcolor[HTML]{E7F1FF}{2}.58 & \cellcolor[HTML]{E7F1FF}{2}.94 & \cellcolor[HTML]{FAFCFC}{3}.24         & \cellcolor[HTML]{FAFCFC}{3}.19 & \cellcolor[HTML]{FAFCFC}{\uline{3.52}} & \cellcolor[HTML]{FAFCFC}{3}.43         & \cellcolor[HTML]{FAFCFC}{3}.39         & \cellcolor[HTML]{FAFCFC}{3}.43         \\
                                 & \lm-7b    & \cellcolor[HTML]{BFDBFF}{1.13}         & \cellcolor[HTML]{FAFCFC}{3}.28 & \cellcolor[HTML]{FAFCFC}{3}.51 & \cellcolor[HTML]{FAFCFC}{\uline{3.88}} & \cellcolor[HTML]{FAFCFC}{3}.8  & \cellcolor[HTML]{FAFCFC}{3}.63         & \cellcolor[HTML]{FAFCFC}{3}.54         & \cellcolor[HTML]{FAFCFC}{3}.42         & \cellcolor[HTML]{FAFCFC}{3}.54         \\
                                 & \lm-13b   & \cellcolor[HTML]{D8E9FF}{1.74}         & \cellcolor[HTML]{FAFCFC}{3}.35 & \cellcolor[HTML]{FAFCFC}{3}.91 & \cellcolor[HTML]{ECF4FF}{4.25}         & \cellcolor[HTML]{DCEBFF}{4.23} & \cellcolor[HTML]{D0E4FF}{4.57}         & \cellcolor[HTML]{C9E1FF}{4.82}         & \cellcolor[HTML]{D4E7FF}{4.91}         & \cellcolor[HTML]{CDE3FF}{\uline{5.13}} \\
                                 & \cl-7b    & \cellcolor[HTML]{FAFCFC}{1.81}         & \cellcolor[HTML]{FAEAE8}{5.78} & \cellcolor[HTML]{F8FBFF}{5.17} & \cellcolor[HTML]{FACFCA}{\uline{5.86}} & \cellcolor[HTML]{FAE0DC}{5.73} & \cellcolor[HTML]{FAE0DC}{5.73}         & \cellcolor[HTML]{ECF4FF}{5.74}         & \cellcolor[HTML]{D6E8FF}{5.49}         & \cellcolor[HTML]{D9EAFF}{5.47}         \\
        \multirow{-7}{*}{Java}   & \cl-13b   & \cellcolor[HTML]{C6DFFF}{1.69}         & \cellcolor[HTML]{FAFCFC}{4.47} & \cellcolor[HTML]{FAE2DF}{4.08} & \cellcolor[HTML]{FAF0EE}{4.14}         & \cellcolor[HTML]{FAE1DE}{4.34} & \cellcolor[HTML]{FADEDA}{4.49}         & \cellcolor[HTML]{FABCB5}{\uline{5.04}} & \cellcolor[HTML]{E9F3FF}{4.61}         & \cellcolor[HTML]{E4F0FF}{4.28}         \\ \midrule
                                 & \ic-1b    & \cellcolor[HTML]{DEECFF}{\uline{4.54}} & \cellcolor[HTML]{FAFCFC}{3}.93 & \cellcolor[HTML]{DCEBFF}{4.23} & \cellcolor[HTML]{EFF6FF}{4.01}         & \cellcolor[HTML]{FAFCFC}{3}.9  & \cellcolor[HTML]{FAFCFC}{3}.88         & \cellcolor[HTML]{FAFCFC}{3}.94         & \cellcolor[HTML]{FAF7F6}{4.07}         & \cellcolor[HTML]{FAFCFC}{3}.92         \\
                                 & \cg-2b-nl & \cellcolor[HTML]{E7F1FF}{2}.99         & \cellcolor[HTML]{FAFCFC}{3}.09 & \cellcolor[HTML]{FAFCFC}{3}.01 & \cellcolor[HTML]{FAFCFC}{3}.26         & \cellcolor[HTML]{FAFCFC}{3}.28 & \cellcolor[HTML]{FAFCFC}{3}.48         & \cellcolor[HTML]{FAFCFC}{3}.55         & \cellcolor[HTML]{FAFCFC}{\uline{3.57}} & \cellcolor[HTML]{FAFCFC}{3}.38         \\
                                 & \cg-6b-nl & \cellcolor[HTML]{FAE2DF}{4.08}         & \cellcolor[HTML]{FAF0EF}{4.35} & \cellcolor[HTML]{EAF3FF}{4.36} & \cellcolor[HTML]{FAE6E3}{4.71}         & \cellcolor[HTML]{FAE6E3}{4.62} & \cellcolor[HTML]{FAFCFC}{4.47}         & \cellcolor[HTML]{FAE3E0}{4.55}         & \cellcolor[HTML]{FAD6D2}{\uline{4.81}} & \cellcolor[HTML]{FAD9D6}{4.77}         \\
                                 & \lm-7b    & \cellcolor[HTML]{E7F1FF}{2}.12         & \cellcolor[HTML]{FAE7E4}{4.68} & \cellcolor[HTML]{FAE1DD}{4.93} & \cellcolor[HTML]{E8F2FF}{5.19}         & \cellcolor[HTML]{C8E0FF}{5.24} & \cellcolor[HTML]{C2DDFF}{\uline{5.31}} & \cellcolor[HTML]{B7D6FF}{5.05}         & \cellcolor[HTML]{B9D7FF}{4.83}         & \cellcolor[HTML]{9EC8FF}{4.64}         \\
                                 & \lm-13b   & \cellcolor[HTML]{FAFCFC}{3}.45         & \cellcolor[HTML]{E1EEFF}{4.6}  & \cellcolor[HTML]{DDECFF}{5.56} & \cellcolor[HTML]{F2F7FF}{6.53}         & \cellcolor[HTML]{E4F0FF}{6.71} & \cellcolor[HTML]{DCEBFF}{\uline{6.75}} & \cellcolor[HTML]{BFDBFF}{6.41}         & \cellcolor[HTML]{A2CBFF}{5.67}         & \cellcolor[HTML]{89BDFF}{5.57}         \\
                                 & \cl-7b    & \cellcolor[HTML]{FEFEFF}{1.5}          & \cellcolor[HTML]{DDECFF}{5.6}  & \cellcolor[HTML]{F0F6FF}{7.03} & \cellcolor[HTML]{EBF4FF}{7.45}         & \cellcolor[HTML]{EEF5FF}{8.07} & \cellcolor[HTML]{FAF1F0}{\uline{8.54}} & \cellcolor[HTML]{F8FBFF}{8.02}         & \cellcolor[HTML]{C4DEFF}{7.16}         & \cellcolor[HTML]{AACFFF}{6.6}          \\
        \multirow{-7}{*}{Python} & \cl-13b   & \cellcolor[HTML]{E7F1FF}{2}.31         & \cellcolor[HTML]{EAF3FF}{4.9}  & \cellcolor[HTML]{FDFEFF}{6.39} & \cellcolor[HTML]{DCEBFF}{6.75}         & \cellcolor[HTML]{BFDBFF}{6.59} & \cellcolor[HTML]{EDF5FF}{\uline{7.28}} & \cellcolor[HTML]{AACFFF}{6.6}2         & \cellcolor[HTML]{DDECFF}{5.56}         & \cellcolor[HTML]{D2E6FF}{5.52}         \\ \bottomrule
    \end{tabular}
    }
\end{table}

\begin{table}[htbp]
\centering
    \caption{{[}RQ3{]} - GLEU and ACC scores of LLM-ICLs with the code input format on JITCU. Blue indicates worse performance while red indicates better performance, compared to using diff as input. The deeper the color, the greater the difference.}
    \label{table_code_icl_jitcu}
    \resizebox{\linewidth}{!}{
        \begin{tabular}{lcccccccccc}
            \toprule
            \textbf{Model} & \multicolumn{1}{l}{\textbf{metric}} & 0shot                                  & 1shot                          & 2shot                                  & 3shot                                   & 4shot                                  & 5shot                           & 6shot                                   & 7shot                           & 8shot                           \\ \midrule
            \ic-1b         &                                     & \cellcolor[HTML]{FADFDC}{\uline{9.99}} & \cellcolor[HTML]{E3EFFF}{0.12} & \cellcolor[HTML]{CDE3FF}{0.47}         & \cellcolor[HTML]{BFDBFF}{1.15}          & \cellcolor[HTML]{C3DDFF}{1.62}         & \cellcolor[HTML]{C2DCFF}{1.65}  & \cellcolor[HTML]{C7DFFF}{2.00}          & \cellcolor[HTML]{C7E0FF}{2.44}  & \cellcolor[HTML]{DCEBFF}{3.19}  \\
            \cg-2b-nl      &                                     & \cellcolor[HTML]{FFFFFF}{0.00}         & \cellcolor[HTML]{FEFEFF}{0.02} & \cellcolor[HTML]{FAFEFE}{1.19}         & \cellcolor[HTML]{FAF6F5}{2.10}          & \cellcolor[HTML]{FAF1F0}{2.91}         & \cellcolor[HTML]{FAE4E1}{4.80}  & \cellcolor[HTML]{FADFDB}{\uline{5.80}}  & \cellcolor[HTML]{FAE1DE}{5.24}  & \cellcolor[HTML]{FAE5E2}{4.74}  \\
            \cg-6b-nl      &                                     & \cellcolor[HTML]{FFFFFF}{0.00}         & \cellcolor[HTML]{FAE9E7}{5.02} & \cellcolor[HTML]{FABCB5}{13.02}        & \cellcolor[HTML]{FAFCFC}{\uline{13.45}} & \cellcolor[HTML]{FCFDFF}{12.00}        & \cellcolor[HTML]{EEF5FF}{8.47}  & \cellcolor[HTML]{E9F3FF}{7.17}          & \cellcolor[HTML]{E8F2FF}{7.69}  & \cellcolor[HTML]{E8F2FF}{7.69}  \\
            \lm-7b         &                                     & \cellcolor[HTML]{F4F9FF}{0.32}         & \cellcolor[HTML]{8CBFFF}{1.68} & \cellcolor[HTML]{91C1FF}{8.02}         & \cellcolor[HTML]{B7D7FF}{13.50}         & \cellcolor[HTML]{B7D7FF}{13.24}        & \cellcolor[HTML]{B9D8FF}{14.83} & \cellcolor[HTML]{C7DFFF}{\uline{15.00}} & \cellcolor[HTML]{B9D8FF}{14.83} & \cellcolor[HTML]{C7E0FF}{14.44} \\
            \lm-13b        &                                     & \cellcolor[HTML]{FCFDFF}{7.50}         & \cellcolor[HTML]{8EC0FF}{4.17} & \cellcolor[HTML]{9BC7FF}{15.29}        & \cellcolor[HTML]{BAD8FF}{\uline{17.49}} & \cellcolor[HTML]{B6D6FF}{15.64}        & \cellcolor[HTML]{BAD8FF}{16.44} & \cellcolor[HTML]{C5DEFF}{16.23}         & \cellcolor[HTML]{C3DDFF}{15.05} & \cellcolor[HTML]{C2DDFF}{14.80} \\
            \cl-7b         &                                     & \cellcolor[HTML]{FAD2CE}{4.89}         & \cellcolor[HTML]{89BDFF}{5.21} & \cellcolor[HTML]{8ABDFF}{9.11}         & \cellcolor[HTML]{ACD0FF}{11.82}         & \cellcolor[HTML]{B2D3FF}{14.53}        & \cellcolor[HTML]{9BC7FF}{15.29} & \cellcolor[HTML]{BDDAFF}{\uline{15.59}} & \cellcolor[HTML]{BDDAFF}{14.07} & \cellcolor[HTML]{C0DCFF}{13.98} \\
            \cl-13b        & \multirow{-7}{*}{GLEU}              & \cellcolor[HTML]{FAEDEB}{2.57}         & \cellcolor[HTML]{94C3FF}{6.15} & \cellcolor[HTML]{B1D3FF}{21.32}        & \cellcolor[HTML]{D5E7FF}{26.07}         & \cellcolor[HTML]{D8E9FF}{26.16}        & \cellcolor[HTML]{D5E7FF}{27.31} & \cellcolor[HTML]{E4F0FF}{\uline{28.48}} & \cellcolor[HTML]{E4F0FF}{26.59} & \cellcolor[HTML]{E7F1FF}{25.84} \\ \midrule
            \ic-1b         &                                     & \cellcolor[HTML]{FAF6F5}{\uline{1.50}} & \cellcolor[HTML]{FAFCFF}{0.05} & \cellcolor[HTML]{FFFFFF}{0.00}         & \cellcolor[HTML]{FAFCFF}{0.05}          & \cellcolor[HTML]{F6FAFF}{0.10}         & \cellcolor[HTML]{FAFCFF}{0.05}  & \cellcolor[HTML]{F5F9FF}{0.15}          & \cellcolor[HTML]{F5F9FF}{0.20}  & \cellcolor[HTML]{F9FBFF}{0.35}  \\
            \cg-2b-nl      &                                     & \cellcolor[HTML]{FFFFFF}{0.00}         & \cellcolor[HTML]{FAFCFF}{0.05} & \cellcolor[HTML]{F5F9FF}{0.20}         & \cellcolor[HTML]{FAFBFA}{\uline{0.60}}  & \cellcolor[HTML]{FAFDFD}{0.30}         & \cellcolor[HTML]{F9FBFF}{0.35}  & \cellcolor[HTML]{FEFEFF}{0.25}          & \cellcolor[HTML]{F5F9FF}{0.15}  & \cellcolor[HTML]{F6FAFF}{0.10}  \\
            \cg-6b-nl      &                                     & \cellcolor[HTML]{FFFFFF}{0.00}         & \cellcolor[HTML]{F9FBFF}{0.35} & \cellcolor[HTML]{FAFCFB}{0.70}         & \cellcolor[HTML]{FDFEFF}{\uline{0.95}}  & \cellcolor[HTML]{FCFDFF}{0.65}         & \cellcolor[HTML]{F5F9FF}{0.20}  & \cellcolor[HTML]{FAFDFD}{0.30}          & \cellcolor[HTML]{F5F9FF}{0.20}  & \cellcolor[HTML]{FEFEFF}{0.25}  \\
            \lm-7b         &                                     & \cellcolor[HTML]{FEFEFF}{0.25}         & \cellcolor[HTML]{DFEDFF}{0.55} & \cellcolor[HTML]{DAEAFF}{1.90}         & \cellcolor[HTML]{E5F0FF}{\uline{2.55}}  & \cellcolor[HTML]{E5F0FF}{2.05}         & \cellcolor[HTML]{E5F0FF}{2.05}  & \cellcolor[HTML]{C7DFFF}{2.00}          & \cellcolor[HTML]{F1F7FF}{1.35}  & \cellcolor[HTML]{EFF6FF}{1.40}  \\
            \lm-13b        &                                     & \cellcolor[HTML]{FDFEFF}{2.60}         & \cellcolor[HTML]{DCEBFF}{1.20} & \cellcolor[HTML]{DAEAFF}{\uline{4.45}} & \cellcolor[HTML]{E1EEFF}{4.35}          & \cellcolor[HTML]{DEECFF}{2.75}         & \cellcolor[HTML]{DFEDFF}{2.25}  & \cellcolor[HTML]{E7F1FF}{1.80}          & \cellcolor[HTML]{FDFEFF}{0.95}  & \cellcolor[HTML]{E9F2FF}{1.25}  \\
            \cl-7b         &                                     & \cellcolor[HTML]{FFFFFF}{0.00}         & \cellcolor[HTML]{D1E5FF}{1.75} & \cellcolor[HTML]{CCE2FF}{1.60}         & \cellcolor[HTML]{E5F0FF}{2.55}          & \cellcolor[HTML]{D9EAFF}{\uline{3.70}} & \cellcolor[HTML]{DAEAFF}{3.55}  & \cellcolor[HTML]{E3EFFF}{3.10}          & \cellcolor[HTML]{C7DFFF}{2.00}  & \cellcolor[HTML]{D1E5FF}{1.75}  \\
            \cl-13b        & \multirow{-7}{*}{ACC}               & \cellcolor[HTML]{DFEDFF}{0.55}         & \cellcolor[HTML]{DFEDFF}{2.25} & \cellcolor[HTML]{DCEBFF}{4.70}         & \cellcolor[HTML]{E8F2FF}{\uline{6.15}}  & \cellcolor[HTML]{E7F1FF}{5.70}         & \cellcolor[HTML]{E5F0FF}{5.90}  & \cellcolor[HTML]{F4F9FF}{5.85}          & \cellcolor[HTML]{F3F8FF}{4.75}  & \cellcolor[HTML]{DAEAFF}{4.45}  \\ \bottomrule
        \end{tabular}
    }
\end{table}

\begin{table}[!htbp]
    \caption{{[}RQ3{]} - Performance of LLM-PEFTs with the code input format on CRG and JITCU. Blue indicates worse performance while red indicates better performance, compared to using diff as input. The deeper the color, the greater the difference.}
    \label{table_code_peft_crg_jitcu}
    \centering
    \resizebox{0.8\textwidth}{!}{
    \begin{tabular}{lcc|cccc}
        \toprule
        \multicolumn{1}{c}{}                                 & \multicolumn{2}{c|}{\textbf{CRG}}    & \multicolumn{4}{c}{\textbf{JITCU}}                                                                                                                                        \\ \cmidrule(l){2-7}
        \multicolumn{1}{c}{}                                 & \multicolumn{2}{c|}{\textbf{BLEU-4}} & \multicolumn{2}{c}{\textbf{GLEU}}  & \multicolumn{2}{c}{\textbf{ACC}}                                                                                                     \\ \cmidrule(l){2-7}
        \multicolumn{1}{c}{\multirow{-3}{*}{\textbf{Model}}} & \textbf{\lora}                       & \textbf{Prefix}                    & \textbf{\lora}                   & \textbf{Prefix}                & \textbf{\lora}                  & \textbf{Prefix}                \\ \midrule
        \ic-1b                                               & \cellcolor[HTML]{FABCB5}{3.54}       & -                                  & \cellcolor[HTML]{89BDFF}{4.98}   & -                              & \cellcolor[HTML]{F3F8FF}{8.85}  & -                              \\
        \cg-2b-nl                                            & \cellcolor[HTML]{FAF2F0}{0.46}       & \cellcolor[HTML]{FAEEED}{0.86}     & \cellcolor[HTML]{FAFBFB}{0.30}   & \cellcolor[HTML]{FBFCFF}{0.03} & \cellcolor[HTML]{FFFFFF}{0.55}  & \cellcolor[HTML]{FABCB5}{0.70} \\
        \cg-6b-nl                                            & \cellcolor[HTML]{FADCD8}{1.65}       & \cellcolor[HTML]{FAF7F7}{0.98}     & \cellcolor[HTML]{FBFDFF}{3.02}   & \cellcolor[HTML]{B1D3FF}{7.55} & \cellcolor[HTML]{FAEAE8}{1.35}  & \cellcolor[HTML]{F8FBFF}{1.65} \\
        \lm-7b                                               & \cellcolor[HTML]{FDFDFF}{5.71}       & \cellcolor[HTML]{FAC5BE}{2.02}     & \cellcolor[HTML]{FCFDFF}{62.58}  & \cellcolor[HTML]{FEFEFF}{0.13} & \cellcolor[HTML]{FAF4F3}{32.35} & \cellcolor[HTML]{FFFFFF}{0.00} \\
        \lm-13b                                              & \cellcolor[HTML]{F6FAFF}{5.16}       & \cellcolor[HTML]{89BDFF}{0.61}     & \cellcolor[HTML]{FAC7C1}{62.35}  & \cellcolor[HTML]{FFFFFF}{0.00} & \cellcolor[HTML]{FAEAE8}{32.65} & \cellcolor[HTML]{FFFFFF}{0.00} \\
        \cl-7b                                               & \cellcolor[HTML]{FAE7E5}{4.40}       & \cellcolor[HTML]{FAF1F0}{0.54}     & \cellcolor[HTML]{F9FBFF}{63.69}  & \cellcolor[HTML]{FAFBFB}{0.09} & \cellcolor[HTML]{FAF9F9}{34.45} & \cellcolor[HTML]{FFFFFF}{0.00} \\
        \cl-13b                                              & \cellcolor[HTML]{F1F7FF}{5.40}       & \cellcolor[HTML]{FCFDFF}{1.13}     & \cellcolor[HTML]{FAFCFF}{63.09}  & \cellcolor[HTML]{FEFEFF}{0.16} & \cellcolor[HTML]{FAF9F9}{34.95} & \cellcolor[HTML]{FFFFFF}{0.00} \\ \bottomrule
    \end{tabular}
    }
\end{table}

\begin{table}[htbp]
    \caption{{[}RQ3{]} - Performance of LLM-PEFTs with the code input format on CMG. Blue indicates worse performance while red indicates better performance, compared to using diff as input. The deeper the color, the greater the difference.}
    \label{table_code_peft_cmg}
    \centering
    \resizebox{\textwidth}{!}{
        \begin{tabular}{lcccccccccc}
            \toprule
            \multicolumn{1}{c}{}                                 & \multicolumn{2}{c}{\textbf{CPP}} & \multicolumn{2}{c}{\textbf{C\#}} & \multicolumn{2}{c}{\textbf{Java}} & \multicolumn{2}{c}{\textbf{JavaScript}} & \multicolumn{2}{c}{\textbf{Python}}                                                                                                                                                                        \\ \cmidrule(l){2-11}
            \multicolumn{1}{c}{\multirow{-2}{*}{\textbf{Model}}} & \textbf{\lora}                    & \textbf{Prefix}                  & \textbf{\lora}                     & \textbf{Prefix}                         & \textbf{\lora}                       & \textbf{Prefix}                & \textbf{\lora}                   & \textbf{Prefix}                & \textbf{\lora}                   & \textbf{Prefix}                \\ \midrule
            \ic-1b                                               & \cellcolor[HTML]{F1F7FF}{5.24}   & -                                & \cellcolor[HTML]{FAFAF9}{5.11}    & -                                       & \cellcolor[HTML]{EEF5FF}{4.75}      & -                              & \cellcolor[HTML]{DBEAFF}{4.66}  & -                              & \cellcolor[HTML]{EBF3FF}{5.12}  & -                              \\
            \cg-2b-nl                                            & \cellcolor[HTML]{FAEDEB}{3.99}   & \cellcolor[HTML]{F9FBFF}{3.47}   & \cellcolor[HTML]{FAEFEE}{3.80}    & \cellcolor[HTML]{EDF5FF}{2.31}          & \cellcolor[HTML]{FAEAE8}{2.95}      & \cellcolor[HTML]{FAECEA}{2.52} & \cellcolor[HTML]{FAFBFA}{3.69}  & \cellcolor[HTML]{CAE1FF}{2.70} & \cellcolor[HTML]{FACDC8}{5.52}  & \cellcolor[HTML]{FAE9E6}{4.29} \\
            \cg-6b-nl                                            & \cellcolor[HTML]{E8F2FF}{4.43}   & \cellcolor[HTML]{E4EFFF}{2.47}   & \cellcolor[HTML]{F1F7FF}{4.47}    & \cellcolor[HTML]{FACBC5}{4.45}          & \cellcolor[HTML]{FAFBFB}{3.77}      & \cellcolor[HTML]{E4EFFF}{2.47} & \cellcolor[HTML]{C6DFFF}{3.87}  & \cellcolor[HTML]{F1F7FF}{3.29} & \cellcolor[HTML]{EEF5FF}{4.90}  & \cellcolor[HTML]{DFEDFF}{4.09} \\
            \lm-7b                                               & \cellcolor[HTML]{FDFEFF}{10.98}  & \cellcolor[HTML]{FABCB5}{2.49}   & \cellcolor[HTML]{FAFCFF}{11.86}   & \cellcolor[HTML]{FAF8F7}{1.28}          & \cellcolor[HTML]{FDFEFF}{12.27}     & \cellcolor[HTML]{F4F9FF}{1.54} & \cellcolor[HTML]{FCFDFF}{13.78} & \cellcolor[HTML]{FAFBFB}{0.15} & \cellcolor[HTML]{EBF3FF}{12.94} & \cellcolor[HTML]{FAE7E5}{0.87} \\
            \lm-13b                                              & \cellcolor[HTML]{FAFCFB}{10.41}  & \cellcolor[HTML]{FAFCFF}{0.59}   & \cellcolor[HTML]{FAFDFD}{11.62}   & \cellcolor[HTML]{FCFDFF}{0.94}          & \cellcolor[HTML]{FAF7F6}{12.01}     & \cellcolor[HTML]{FAFBFB}{0.83} & \cellcolor[HTML]{FDFEFF}{13.36} & \cellcolor[HTML]{FAF8F7}{1.27} & \cellcolor[HTML]{FBFDFF}{12.78} & \cellcolor[HTML]{FBFDFF}{0.77} \\
            \cl-7b                                               & \cellcolor[HTML]{F6FAFF}{10.82}  & \cellcolor[HTML]{FAEFED}{1.44}   & \cellcolor[HTML]{FBFDFF}{11.98}   & \cellcolor[HTML]{DCEBFF}{0.41}          & \cellcolor[HTML]{FAFDFD}{13.10}     & \cellcolor[HTML]{FAF5F4}{1.45} & \cellcolor[HTML]{FBFDFF}{13.72} & \cellcolor[HTML]{FACBC5}{2.67} & \cellcolor[HTML]{FAFBFB}{13.44} & \cellcolor[HTML]{FAFDFD}{0.65} \\
            \cl-13b                                              & \cellcolor[HTML]{ECF4FF}{11.36}  & \cellcolor[HTML]{C5DEFF}{1.20}   & \cellcolor[HTML]{FEFEFF}{12.10}   & \cellcolor[HTML]{C8E0FF}{1.74}          & \cellcolor[HTML]{FEFEFF}{12.10}     & \cellcolor[HTML]{FAD2CD}{3.26} & \cellcolor[HTML]{FAFBFB}{13.44} & \cellcolor[HTML]{FAE1DE}{2.63} & \cellcolor[HTML]{BCD9FF}{11.42} & \cellcolor[HTML]{89BDFF}{1.26} \\ \bottomrule
        \end{tabular}
    }
\end{table}

Table \ref{table_code_icl_crg}, \ref{table_code_icl_cmg}, \ref{table_code_icl_jitcu} and Table \ref{table_code_peft_crg_jitcu}, \ref{table_code_peft_cmg} show the performance \icl s and \peft s using code format as input.
To visually demonstrate the performance difference between using code as input and using diff as input, we use blue to indicate a relatively worse performance and red to indicate a relatively better performance.
The deeper the color, the greater the difference.

\subsubsection{Impact of Different Input Formats on LLM-ICL}

We observe that when using LLM-ICL, using diff as the input of LLMs generally outperforms using code as the input of LLMs.
Specifically, on JITCU, the performance of LLMs is significantly better when using diff as input compared to using code as input (except for the \cg~family, where the ACC value is close to 0 across different settings).
For example, on CRG, the best-performing LLM-ICL (\cl-7b) has a BLEU score of 5.04 when using diff as input, which is only 0.14 lower than the best result obtained using code as input.
Additionally, for the best-performing LLM-ICL on JITCU (\cl-7B), the ACC score is 23.65 when using diff as input, while it is only 1.6 when using code as input.
One possible reason is that how to update the comment is highly related to the changed parts in the code. Diffs explicitly annotate the changed lines in code with ``+" an ``-", making it easier for LLMs to identify and understand what is changed.
For two connected code snippets, LLM needs to first understand the two snippets, then determine the updated, deleted, and unchanged lines of code by comparing them, and finally comprehend the change information, which is more complex.

\rqbox{\textbf{Finding 8}:
    When using LLM-ICL, the input format significantly affects the performance of LLM. Specifically, the model performs better with diff as input, especially on the JITCU task.
}\label{finding8}

\subsubsection{Impact of Different Input Formats on LLM-PEFT}

For LLM-PEFTs, there is no significant performance difference between different input formats.
For example, when using LLM-PEFT, on CRG, the performance difference between different input formats is only 0.03; on JITCU, the differences in GLEU and ACC are only 1.54 and 0.05, respectively.
These facts indicate that LLM-PEFTs are not sensitive to the input format, which is different from the LLM-ICLs.
One possible reason is that when representing a code change as two code snippets, it is difficult for LLMs without fine-tuning to compare two code snippets and capture the changed part, while LLMs fine-tuned with some code change data can learn to adapt to the input format and thus achieve better performance.

\rqbox{\textbf{Finding 9}:
    When using LLM-PEFT, the input format does not significantly affect the performance of the LLM.
}\label{finding9}

\subsection{RQ5: When Do LLMs Perform Better?}\label{rq_content}

\begin{table}[!htbp]
    \caption{{[}RQ5{]} - The selected pre-trained models for different tasks under the application of ICL or Finetuning.}
    \label{table_content_models}
    \centering
    \resizebox{0.8\linewidth}{!}{
    \begin{tabular}{lm{3cm}<{\centering}ll}
        \toprule
                                                 & \textbf{fully fine-tuned small pre-trained model} & \textbf{LLM-ICL}                 & \textbf{LLM-PEFT}       \\
        \midrule
        \textbf{CRG}                             & \multirow{7}{*}{\cc}                             & \multirow{7}{*}{\cl-7b} & \lm-7b              \\
        \textbf{CMG\_\scalebox{0.8}{Java}}       &                                                   &                                  & \cl-7b         \\
        \textbf{CMG\_\scalebox{0.8}{C\#}}        &                                                   &                                  & \cl-13b        \\
        \textbf{CMG\_\scalebox{0.8}{Cpp}}        &                                                   &                                  & \cl-13b        \\
        \textbf{CMG\_\scalebox{0.8}{Python}}     &                                                   &                                  & \cl-7b         \\
        \textbf{CMG\_\scalebox{0.8}{JavaScript}} &                                                   &                                  & \cl-13b        \\
        \textbf{JITCU}                           &                                                   &                                  & \cl-7b \\
        \bottomrule
    \end{tabular}
    }
\end{table}

\begin{table}[!htbp]
    \caption{{[}RQ5{]} - Performance of different techniques for each code change category in code-change-related tasks. BLEU for CRG, CMG, GLEU for JITCU. The deeper the color, the better the performance.}
    \label{table_content_result}
    \centering
    \resizebox{0.75\linewidth}{!}{
    \begin{tabular}{clccccc}
        \toprule
                                &                          &                                 & \multicolumn{2}{c}{Code}        &                                 &                                                                   \\ \cmidrule(lr){4-5}
        \multirow{-2}{*}{Task}  & \multirow{-2}{*}{Method} & \multirow{-2}{*}{Doc}           & Feat                            & Ref                             & \multirow{-2}{*}{Doc\&Code}     & \multirow{-2}{*}{Total}         \\ \midrule
                                & Sample Num               & 4                               & 149                             & 13                              & 30                              & 196                             \\ \cmidrule(l){2-7}
                                & Small Pre-trained Models & \cellcolor[HTML]{FF770F}{5.48}  & \cellcolor[HTML]{FF8A30}{4.71}  & \cellcolor[HTML]{FF8222}{5.04}  & \cellcolor[HTML]{FF7811}{5.44}  & \cellcolor[HTML]{FF872B}{4.83}  \\
                                & LLM-ICL                  & \cellcolor[HTML]{FFB176}{3.12}  & \cellcolor[HTML]{FF8B32}{4.68}  & \cellcolor[HTML]{FFA45E}{3.67}  & \cellcolor[HTML]{FFA058}{3.81}  & \cellcolor[HTML]{FF913C}{4.45}  \\
        \multirow{-4}{*}{CRG}   & LLM-PEFT                 & \cellcolor[HTML]{FF6F00}{5.83}  & \cellcolor[HTML]{FF7811}{5.43}  & \cellcolor[HTML]{FFA158}{3.80}  & \cellcolor[HTML]{FF8323}{5.02}  & \cellcolor[HTML]{FF7C18}{5.27}  \\ \midrule
                                & Sample Num               & 12                              & 79                              & 56                              & 53                              & 200                             \\ \cmidrule(l){2-7}
                                & Small Pre-trained Models & \cellcolor[HTML]{FFBD8B}{9.86}  & \cellcolor[HTML]{FFB176}{11.65} & \cellcolor[HTML]{FF6F00}{21.71} & \cellcolor[HTML]{FFA25B}{13.94} & \cellcolor[HTML]{FF994B}{15.31} \\
                                & LLM-ICL                  & \cellcolor[HTML]{FFBE8D}{9.68}  & \cellcolor[HTML]{FFD5B5}{6.30}  & \cellcolor[HTML]{FFBD8B}{9.82}  & \cellcolor[HTML]{FFDBBF}{5.37}  & \cellcolor[HTML]{FFD3B1}{6.62}  \\
        \multirow{-4}{*}{CMG}   & LLM-PEFT                 & \cellcolor[HTML]{FF8629}{18.19} & \cellcolor[HTML]{FFBC89}{10.04} & \cellcolor[HTML]{FF801E}{19.13} & \cellcolor[HTML]{FFA057}{14.30} & \cellcolor[HTML]{FFA866}{13.01} \\ \midrule
                                & Sample Num               & 0                               & 102                             & 81                              & 13                              & \textbf{196}                    \\ \cmidrule(l){2-7}
                                & LLM-ICL                  & -                               & \cellcolor[HTML]{FF984A}{53.58} & \cellcolor[HTML]{FF770E}{71.17} & \cellcolor[HTML]{FFBD8B}{34.14} & \cellcolor[HTML]{FF8C34}{59.97} \\
                                & LLM-PEFT                 & -                               & \cellcolor[HTML]{FF913D}{57.42} & \cellcolor[HTML]{FF6F00}{75.57} & \cellcolor[HTML]{FFAA69}{44.45} & \cellcolor[HTML]{FF8426}{64.13} \\
        \multirow{-4}{*}{JITCU} & Small Pre-trained Models & -                               & \cellcolor[HTML]{FF8F39}{58.43} & \cellcolor[HTML]{FF740A}{72.59} & \cellcolor[HTML]{FFA662}{46.38} & \cellcolor[HTML]{FF8222}{65.35} \\ \bottomrule
    \end{tabular}
}
\end{table}

Table \ref{table_content_models} shows the best-performing models from small pre-trained models, LLM-ICLs and LLM-PEFTs selected to answer this research question.
Table \ref{table_content_result} shows the performance of the models on each code change category for each code-change-related task.

We observe that \textbf{the best-performing LLM-PEFTs outperform the best-performing LLM-ICLs on all types of code changes}
For example, on CRG, the best-performing LLM-PEFTs outperform the best-performing LLM-ICLs on all types of code changes by 86.86\%, 16.03\%, 3.54\%, and 31.76\% on each code change type, respectively.
This indicates that compared to ICL, PEFT can comprehensively improve the performance of LLMs on all code change categories.
Therefore, we believe that LLM-PEFT can be applied to different types of code changes.

\rqbox{\textbf{Finding 13}:
    PEFT can comprehensively improve the performance of LLMs across all categories of code changes, including changes that only modify documentation, only modify code, and those that modify both code and documentation simultaneously.
}\label{finding13}

We observe that \textbf{the best-performing LLM-PEFTs statistically significantly outperform the best-performing fully fine-tuned small models on doc-only code changes.}
Specifically, in terms of doc-only code changes on CRG and CMG, the best-performing LLM-PEFTs achieve the BLEU scores of 5.83 and 18.19, respectively, while those of the best-performing fully fine-tuned small models are 5.48 and 9.86. The results indicate that LLM-PEFTs are effective on doc-only code changes.
The possible reason is that LLMs are good at understanding natural languages.

In terms of the changes related to source code, the best-performing LLM-PEFTs perform comparably to fully fine-tuned small models, indicating LLM-PEFTs can to some extent understand the changes of functional logic.
For example, on JITCU, in terms of code changes related to Ref, the best-performing LLM-PEFTs achieve the BLEU score of 75.57, while that of the best-performing fully fine-tuned small models is 72.59.
on CMG, in terms of code changes related to Feat, the best-performing fully fine-tuned \sm s can achieve the BLEU score of 11.65, while that of the best-performing LLM-PEFTs is 10.04.
This indicates that compared with \sm s, though LLM-PEFTs can understand the changes that are only related to documentation, LLM-PEFTs still need more knowledge to understand featuring, refactoring, and the mix of code and documentation modifications.
One possible reason is that fully fine-tuned small models have learned the knowledge related to source code changes through the pre-training specific to code changes, while LLM-PEFTs are only fine-tuned with some task-specific data.
Thus, LLM-PEFTs may have insufficient code-change-specific knowledge, such as the knowledge of refactoring.
We recommend that future work consider taking into account code-change-oriented data and objectives during the pre-training of LLM for code-change-related tasks.

\rqbox{\textbf{Finding 14}:
    LLM-PEFTs can outperform the fully fine-tuned small models on the doc-only code changes and achieve comparable performance to fully fine-tuned small models on other code change types.
}\label{finding14}

\section{DISCUSSION}

\subsection{Human Evaluation}
\label{dis_human}

Here, following prior studies \cite{hu2022correlating, liu2024non, lin2023cct5}, we take CMG as an example and conduct a human evaluation to further investigate the effectiveness of LLMs in code-change-related tasks.

Specifically, following prior studies \cite{lin2023cct5,tao2022large}, we recruit 4 evaluators who are not the co-authors of this paper.
All evaluators have more than 5 years of software development experience and have a deep understanding of Computer Science.
We randomly sample 100 commits from the MCMD dataset \cite{tao2022large} (50 commits for Java, and 50 commmits for Python).
The number of sampled commits is the same as the number of sampled commits used in prior studies related to the human evaluation of the model performance on the MCMD dataset \cite{lin2023cct5}.
We chose these two languages (i.e., Java and Python) because they are currently the most popular programming languages. Additionally, compared to the other three languages, the evaluators are more familiar with these two languages.
We then apply the best-performing models from the explored three approaches (\sm s, LLM-ICLs, and LLM-PEFTs) to generate commit messages for these sampled commits.
Specifically, in terms of small pre-trained models, we apply \cc~with the diff input.
In terms of LLM-ICL, we apply \cl-7b with diff input using 7 examples on the Java sub-dataset, and \cl-7b with diff input using 5 examples on the Python sub-dataset.
For LLM-PEFT, we apply \cl-7b-\lora~with diff input on Java and Python sub-datasets.
This process yields 300 generated commit messages.

Following prior studies~\cite{lin2023cct5,tao2022large}, we ask evaluators to rate each generated message from the following three dimensions, with each dimension scored on a scale of 1 to 5 (1: poor, 2: marginal, 3: acceptable, 4: good, 5: excellent):

\begin{enumerate}
	\item \textbf{Adequacy}: The extent to which the generated commit message covers the main content of the code changes. A higher score indicates that the generated information more comprehensively summarizes the main content of the code modifications.
	\item \textbf{Conciseness}: The extent to which the generated commit message does not contain irrelevant content. A higher score indicates that the information is more concise, without the interference of redundant and irrelevant information.
	\item \textbf{Expressiveness}: The readability and understandability of the generated commit message. A higher score indicates that the information is expressed more clearly and easier to understand.
\end{enumerate}

To facilitate the evaluation process, we prepare a questionnaire for each commit. The questionnaire includes the code change, the ground truth commit message, and the commit messages generated by the three compared techniques (\sm s, LLM-ICLs, and LLM-PEFTs). To minimize potential bias, the three techniques are anonymous in the questionnaire, ensuring that the participants are unaware of which technique generated each commit message. Furthermore, each participant completes the questionnaire independently, without any discussion or collaboration with others.
Each sample commit will be evaluated by two participants, and we consider the average of the two scores as the final score. If the two scores are significantly different, we will invite a third evaluator to evaluate the commit, and the final score will be the average of the two closest scores.

\begin{table}[!htbp]
    \caption{{[}Discussion{]} the results of our human evaluations}
    \label{table_human_result}
    \begin{tabular}{l|ccc}
        \hline
        \textbf{Approach}       & \multicolumn{1}{l}{\textbf{Adequacy}} & \multicolumn{1}{l}{\textbf{Conciseness}} & \multicolumn{1}{l}{\textbf{Expressiveness}} \\ \hline
        LLM-ICL                 & 2.46                                  & 2.72                                     & 3.82                                        \\
        Small Pre-trained Model & \textbf{3.41}                         & 3.66                                     & 4                                           \\
        LLM-PEFT                & 3.29                                  & \textbf{3.78}                            & \textbf{4.23}                               \\ \hline
    \end{tabular}
\end{table}

Same as the conclusion in Section \ref{rq_compare}, the human evaluation further confirms that \textbf{LLM-PEFTs can generate commit messages that are more concise and readable compared with other techniques}.
Table~\ref{table_human_result} presents the results of the human evaluation. Overall, the LLM-PEFTs outperform the other techniques in terms of conciseness and expressiveness.
The average scores achieved by LLM-PEFTs for adequacy, conciseness, and expressiveness are 3.29, 3.78, and 4.22, respectively.
These values are 33.74\%, 38.97\% and 10.76\% higher than those of LLM-ICLs, and -3.52\%, 2.92\%, and 5.50\% higher than those of small pre-trained models, respectively.
This indicates that LLM-PEFTs can generate more concise and readable commit messages compared with other techniques.
While the adequacy of LLM-PEFTs is slightly lower than that of LLM-ICLs, the difference is not significant.
This indicates that LLM-PEFTs can generate comparable commit messages compared with \sm s in terms of adequacy.

Interestingly, we also observe that LLM-PEFTs are rated higher than the fully fine-tuned small models in terms of conciseness.
This is counter-intuition as it is generally considered that LLMs can generate informative and readable texts, but often with some redundant information~\cite{imran2024uncovering,liu2016not}.
One possible reason is that through fine-tuning, LLMs have learned the preference for this task, i.e., to generate concise summaries rather than lengthy words.

\begin{table}[!htbp]
  \caption{{[}Discussion{]} - An example of the commit message generation task.}
  \centering
  \label{table_human_example}
  \small
  \begin{tabular}{l|l}
    \hline
\textbf{Diff}       & \begin{tabular}[c]{@{}l@{}}
\begin{lstlisting}[language=python]
exports.update_messages=function update_messages(events){
}
msgs_to_rerender.push(msg);

- msg.alerted=event.flags.indexOf("has_alert_word")!==-1;
- msg.mentioned=event.flags.indexOf("mentioned")!==-1||
- event.flags.indexOf("wildcard_mentioned")!==-1;
+ message_store.set_message_booleans(msg,event.flags);

condense.un_cache_message_content_height(msg.id);
\end{lstlisting}
 \end{tabular}                            \\ \hline

    \textbf{Gold}                          & Call message.set\_message\_booleans() in update\_messages().                   \\ \hline
    \textbf{LLM-ICL}                       & Merge pull request from jeannefukumaru/update\_message\_store                                        \\
    \textbf{Fully Fine-tuned Small Models} & refactor:Use set\_message\_booleans event for mentions. \\
    \textbf{LLM-PEFT}                      & message\_store.set\_message\_booleans()                                              \\ \hline
    
  \end{tabular}
\end{table}

To better understand the quality of the generated commit message, we conduct a case study.
Table~\ref{table_human_example} shows an example of the commit message generation task.
We present the changed code, the commit message written by developers, as well as the commit messages generated by LLM-ICL, fully fine-tuned small models, and LLM-PEFT.
Based on the diff and the commit messages written by the developers, we can see that the code change is related to \textit{refactoring} the \textit{update\_messages()} using the \textit{set\_message\_booleans()} in \textit{message\_store} class.
However, the commit messages generated by the three techniques are not perfect.
The commit message generated by LLM-ICL mentions a plausible phrase \textit{update\_message\_store}, but it is not accurate; all the information is hallucinated, but its readability is acceptable.
This indicates that the commit message generated by LLM-ICL is not adequate and concise at all, but expressive.
The commit message generated by fully fine-tuned small models mentions the \textit{refactor} and \textit{set\_message\_booleans()}, but misses \textit{set\_message\_booleans()} and \textit{message\_store}; it hallucinates the information \textit{event} and \textit{mentions}, but its readability is acceptable.
This indicates that the commit message generated by fully fine-tuned small models is moderate in terms of adequacy and conciseness, and is expressive.
The commit message generated by LLM-PEFT mentions the \textit{set\_message\_booleans()} and \textit{message\_store}, but misses \textit{refactoring} and \textit{update\_messages()}; there is no hallucinated information, and its readability is acceptable.
This indicates that the commit message generated by LLM-PEFT is moderate in terms of adequacy, and is expressive and very concise.

\rqbox{\textbf{Finding 15}:
	Users rate that compared to other techniques, LLM-PEFT can generate concise and readable commit messages without compromising too much adequacy.
}\label{finding15}

\subsection{Implications}

\textbf{When adapting LLMs with ICL to code-change-related tasks, the number of examples in the prompt should be determined by the data length of the task and the context length allocated to the model.}
In Section \ref{rq_icl}, we observe that the performance of LLM-ICLs is not always positively correlated with the number of examples, and the best-performing LLM-ICLs have different numbers of examples in different tasks.
One possible reason is that the lengths of examples in the different tasks are different, and the context length allocated to the model is limited.
Therefore, we suggest that practitioners should make a decision based on the distribution of data lengths in the task and the context length allocated to the model when adapting LLMs with ICL to code-change-related tasks.

\textbf{When using LLMs on code-change-related tasks, the choice of the model family is more important than the model size}.
In Section \ref{rq_icl} and Section \ref{rq_peft}, we observe that the performance of LLM is not always positively correlated with the model size even within the same model family, no matter whether using ICL or PEFT.
However, we observe that there are differences between the performance of LLMs from different model families.
For example, when using LLM with ICL, the best-performing \cl~outperforms the best-performing \lm~and \cg~across tasks.
However, the best-performing smaller \cl~outperforms the best-performing larger \cl across tasks when using LLM with ICL.
This indicates that the choice of the model family is more important than the model size when using LLMs on code-change-related tasks.
Therefore, we suggest that when an LLM is not performing well on code-change-related tasks, practitioners should not only try different model sizes but also try different model families.

\textbf{Code-change-related tasks benefit more from the PEFT techniques that can help LLMs learn new knowledge.}
Section~\ref{rq_peft} shows that LLMs tuned with \lora~outperform the LLMs tuned with prefix-tuning in almost all settings.
One main reason is that \lora~can better help LLMs learn new knowledge, i.e., the knowledge of code changes, while prefix-tuning mainly focuses on stimulating the learned knowledge in LLMs.
Therefore, when applying LLMs to code-change-related tasks, we suggest using the PEFT techniques that can effectively help LLMs learn new knowledge, such as 
\lora.

\textbf{LLM-ICLs open up the opportunities to automate the code-change-related tasks suffering from data scarcity.}
Data scarcity refers to situations where data acquisition is difficult or the amount of data is small.
This often occurs in new tasks or scenarios involving user privacy sensitivities.
Limited training data cannot guide pre-trained models, such as CodeT5 and \cc, to learn downstream tasks well through fine-tuning, resulting in their poor performance.
We have also tried to adapt CodeT5 and \cc~with ICL and no fine-tuning to code-change-related tasks, but find they cannot produce meaningful output.
In contrast, Section \ref{rq_compare} has shown that LLM-ICLs can achieve promising performance on code-change-related tasks with only a few examples due to LLMs' emergent abilities.
These results imply that it is now possible to automate the code-change-related tasks suffering from data scarcity with LLM-ICLs, which we believe can be a promising research direction.

\textbf{Pre-training LLMs with code-change-oriented objectives can potentially bring substantial improvements to code-change-related tasks.}
We can see from Section~\ref{rq_compare} that on CRG and CMG, \cc~outperforms CodeT5 by substantial margins, and achieves comparable performance to the best LLM-PEFTs. On JITCU, the ACC score of \cc~is higher than CodeT5 but lower than the best LLM-PEFTs.
Comparing \cc~with CodeT5, we can see that pre-training models using code-change-oriented objectives are beneficial for code-change-related tasks.
Comparing LLM-PEFTs with CodeT5, we can know that increasing the sizes of the model and the pre-training data are also beneficial.
Inspired by these, a straightforward idea would be pre-training an LLM with code-change-oriented objectives, which can take advantage of the two beneficial designs.
We believe this direction would bring substantial benefits for code-change-related tasks.

\textbf{Practitioners can potentially improve LLM-ICLs for code-change-related tasks by designing novel formats to represent code changes}. Diffs and code snippets are the most common formats to represent code changes. In Section~\ref{rq_input}, we investigate the impact of input formats on the effectiveness of LLM-ICLs, and find that LLM-ICLs with diffs as the input format outperforms those representing code changes as code snippets in most of the settings, especially on the JITCUP tasks. This indicates that the representation formats of code changes play an important role in the effectiveness of LLM-ICLs. Thus, it is interesting and promising to improve LLM-ICLs on code-change-related tasks by designing novel representation formats for code changes. For example, we can identify the changed parts in each code change using static analysis tools and explicitly mention these changed parts in the prompt to help LLMs better understand code changes.

\textbf{To boost code-change-related tasks with LLMs, we should guide LLMs to learn more code-change-specific knowledge.}
In Section \ref{rq_content}, we observe that LLM-PEFTs significantly outperform the fully fine-tuned small models in doc-only code changes.
This is related to the fact that LLMs are pre-trained with rich sources of natural language texts and thus have the ability to understand and handle tasks related to natural languages.
However, the fully fine-tuned small models perform better than LLM-PEFTs on the refactor and doc-and-code code changes.
This indicates that PEFT on task-specific datasets is not enough for LLMs to learn code-change-specific knowledge, such as the knowledge of refactoring.
We suggest that future researchers should focus more on guiding LLMs to learn code-change-specific knowledge when tuning LLMs for code-change-related tasks. For example, we can leverage the idea of multi-task learning and transfer learning. Before fine-tuning LLMs on the downstream task, we can first tune LLMs on other tasks that are helpful for identifying code-change-specific knowledge, such as refactoring detection, and then fine-tune them on the specific downstream tasks.

\subsection{Threats to Validity}\label{sec_threat}
We identify three primary threats to the validity of our study:

1) \textbf{The selection of LMs.} To mitigate this threat, we design specific selection criteria for model choice, as demonstrated in Section \ref{sec:models}. These models include LLMs of different sizes from various families, which are trained with different pre-training data and learning objectives. Additionally, we have also selected robust small pre-trained models, like CCT5, to thoroughly investigate the impact of fine-tuning on LMs.

2) \textbf{Potential data leakage.} In this paper, we conduct experiments on code changes using multiple LMs. However, these models are trained with large amounts of data, which may raise concerns about potential data leakage. It is possible that these models have seen some data in our test sets and merely memorize the results instead of predicting them. Nevertheless, during our exploration of the ICL process, we find that the selected LMs exhibit poor performance in zero-shot scenarios, indicating that there is no evidence of LMs memorizing the dataset. Additionally, in our experiments, we measure different strategies using LLMs through relative performance comparisons. Therefore, we believe the findings of our paper remain valid.

3) \textbf{Representativeness of automated evaluation metrics.}
Though automated evaluation metrics (BLEU-4, ACC, GLEU) are widely used in the field of natural language processing, they may not fully reflect the human perception of the text.
To address this limitation, we take the commit message generation task as an example and conduct a human evaluation.
The evaluation result is consistent with the automated metrics.
We believe that the automated metrics are reliable for evaluating the performance of LLMs on code-change-related tasks.

4) \textbf{Uncertainty caused by manual data annotation.} In Section \ref{rq_compare} and Section \ref{dis_human}, we randomly selected samples based on previous work~\cite{lin2023cct5,tao2022large} and manually labelled them.
Subjectivity in human decisions is a potential threat during the manual labeling process.
To address this threat, following previous studies ~\cite{lin2023cct5,tao2022large}, each sample is labeled by more than one participant.
The disagreement between the participants is resolved through discussion.
We believe such results are reliable.

\section{Related Work}

\subsection{LLM in Software Engineering}

Pre-trained models have demonstrated impressive capabilities in the field of natural language processing and have shown their excellent performance on a wide range of applications in various domains \cite{lin2023cct5, 10336331, li2022automating, zhang2022coditt5,lu2023llama}.
Recently, Large Language Models (LLMs) have been introduced equipped with billions of parameters and billions of training samples \cite{zan-etal-2023-large}.
LLMs are pre-trained on large-scale text data and learn rich linguistic knowledge and semantic representations which enable them to understand the meaning and structure of natural language.
In the field of software engineering, researchers enhanced the general LLMs on code-related tasks and proposed many domain-specific LLMs, e.g., InCoder\cite{fried2022incoder}, Code Llama\cite{roziere2023code}, StarCoder\cite{li2023starcoder}, and SantaCoder\cite{allal2023santacoder}.
A series of studies have experimentally demonstrated that these domain-specific LLMs achieve good performance on code-related tasks, e.g., code completion, automatic code generation, code understanding, and code summarization.
However, these models were only pre-trained on the code-related tasks and ignored the code-change-related tasks.
Pre-training LLM using code-related tasks focuses on the general syntactic and semantic knowledge of code, while code changes are more concerned with the differences between two code snippets.
It is still unclear how these LLMs perform on the code-change-related tasks.
To fill the gap, in our work, we explore the capabilities of representative LLMs on code-change-related tasks and show the promising directions of using LLMs on code-change-related tasks.

\subsection{Technique for Code-Change-Related Task}

Prior studies proposed a series of approaches for code-changes-related tasks.
For example, Jiang et al.\cite{jiang2017automatically} adapted Neural Machine Translation (NMT) to automatically ``translate'' diffs into commit messages.
Liu et al.\cite{liu2020automating} focused on the task of comment update.
They leveraged a sequence-to-sequence model to learn comment update patterns from code-comment co-changes.
Hoang et al.\cite{hoang2020cc2vec} employed the attention mechanism to model the hierarchical structure of a code change, and used multiple comparison functions to identify the differences between the removed and added code.
They evaluated the performance of the proposed approach on log message generation, bug fixing patch identification, and just-in-time defect prediction, and the proposed approach outperformed the state-of-the-art techniques.
Recently, pre-trained models have been proposed for code-change-related tasks.
Researchers pre-trained a large amount of code change data with code-change-oriented objectives \cite{lin2023cct5, 10336331, li2022automating, zhang2022coditt5,lu2023llama}.
For example, Lin et al. \cite{lin2023cct5} proposed CCT5, which trained with CodeT5 with 5 kinds of code-change-oriented objectives.
Zhou et al. \cite{10336331} proposed CCBERT, which was pre-trained on four proposed self-supervised objectives that are specialized for learning code change representations based on the contents of code changes.
Different from these works that only considered the models with millions of parameters, our work explores the capability of code changes on LLMs with more parameters.

The most related work to our paper is the work of Liu et al. \cite{liu2024delving}.
Specifically, Liu et al. investigated using PEFT on small pre-trained models (\textless 1B parameters) on 2 code change-related tasks, i.e., CMG and Just-In-Time Defect Prediction.
Different from their work, we focus on LLM (\textgreater 1B parameters).
Such models have been proposed more recently, and their capability on code change-related tasks has not been systematically explored.
Besides, we focus on more code change-related tasks, i.e., CRG, CMG, and JITCU.
All of these tasks are generation tasks, which are more challenging than the classification task used in Liu et al. \cite{liu2024delving}'s work (i.e., Just-In-Time Defect Prediction).
Furthermore, besides PEFT, we also explore ICL, which shows promising ability with LLMs (\textgreater 1B parameters).
Our results show that on the CRG task, LLMs with ICL can outperform CodeT5 (a small pre-trained code model fine-tuned on code-change-related tasks), and the best-performing LLM-PEFTs have comparable performance to the state-of-the-art fully fine-tuned small models, indicating that LLMs are promising on code-change-related tasks.

\section{Conclusion and Future Work}
In this paper, we conduct an empirical study to explore the capabilities of LLMs on code-change-related tasks. Specifically, we adapt LLMs with in-context learning (ICL) and parameter-efficient fine-tuning (PEFT), respectively, to three code-change-related tasks, i.e., code review generation, commit message generation, and just-in-time comment update. We investigate the effects of multiple factors, such as the number of examples for ICL, and the choice of PEFT methods, and we compare the performance of LLMs with that of \sm s.
We also explore the impact of the format of code changes and the impact of code change categories on the performance of LLMs.
Experimental results show that LLMs are promising for code-change-related tasks, and the best-performing LLMs are often achieved by tuning \lm~or \cl~using \lora~across model sizes.
At the same time, LLMs tend to learn the code change knowledge related to documents.
We summarize our findings and provide better suggestions to help practitioners better adapt LLMs to code-change-related tasks.
In the future, we will try to explore the effects on the performance of LLMs in code-change-related tasks combined with more aspects of code changes, such as the impact of different code change representation forms and the impact of introducing more knowledge related to refactorings to LLMs.

\section*{Data availability}
Our replication package, including the source code and used datasets, is available at:
\url{https://github.com/ZJU-CTAG/CodeChange}.

\bibliographystyle{plain}
\bibliography{paper}

\appendix
\section{Appendix}\label{sec_appendix}

\begin{table*}[h]
\caption{Performance of LLM-ICL on CMG with diff as input.}
\label{tab:append-cmg}
\centering
\resizebox{\textwidth}{!}{
\begin{tabular}{clccccccccc}
\toprule
\multicolumn{1}{l}{Lang} & Model & 0shot & 1shot & 2shot & 3shot & 4shot & 5shot & 6shot & 7shot & 8shot \\ \midrule
 & InCoder-1b & \cellcolor[HTML]{FFC191}{3.59} & \cellcolor[HTML]{FFCCA4}{2.97} & \cellcolor[HTML]{FFCEA9}{2.8} & \cellcolor[HTML]{FFCDA6}{2.91} & \cellcolor[HTML]{FFD2B0}{2.58} & \cellcolor[HTML]{FFCEA9}{2.8}6 & \cellcolor[HTML]{FFCFAA}{2.79} & \cellcolor[HTML]{FFCBA3}{3} & \cellcolor[HTML]{FFD0AC}{2.71} \\
 & CodeGen-2b-nl & \cellcolor[HTML]{FFDEC5}{1.9} & \cellcolor[HTML]{FFD9BB}{2.21} & \cellcolor[HTML]{FFD5B5}{2.41} & \cellcolor[HTML]{FFD3B1}{2.54} & \cellcolor[HTML]{FFD5B5}{2.41} & \cellcolor[HTML]{FFD6B7}{2.34} & \cellcolor[HTML]{FFD9BB}{2.21} & \cellcolor[HTML]{FFDBBF}{2.09} & \cellcolor[HTML]{FFD8BA}{2.26} \\
 & CodeGen-6b-nl & \cellcolor[HTML]{FFDABF}{2.1} & \cellcolor[HTML]{FFCEA9}{2.8}7 & \cellcolor[HTML]{FFCBA3}{3}.28 & \cellcolor[HTML]{FFCBA3}{3}.17 & \cellcolor[HTML]{FFCBA3}{3}.13 & \cellcolor[HTML]{FFCBA3}{3}.07 & \cellcolor[HTML]{FFCBA3}{3}.09 & \cellcolor[HTML]{FFCBA3}{3}.16 & \cellcolor[HTML]{FFCBA3}{3}.11 \\
 & Llama-2-7b & \cellcolor[HTML]{FFD8BB}{2.23} & \cellcolor[HTML]{FFCBA3}{3}.48 & \cellcolor[HTML]{FFCBA3}{3}.63 & \cellcolor[HTML]{FFCBA3}{3}.95 & \cellcolor[HTML]{FFCBA3}{3}.97 & \cellcolor[HTML]{FFB882}{4.08} & \cellcolor[HTML]{FFB780}{4.16} & \cellcolor[HTML]{FFB984}{4.04} & \cellcolor[HTML]{FFB77F}{4.18} \\
 & Llama-2-13b & \cellcolor[HTML]{FFD5B5}{2.42} & \cellcolor[HTML]{FFCBA3}{3}.68 & \cellcolor[HTML]{FFCBA3}{3}.8 & \cellcolor[HTML]{FFB073}{4.58} & \cellcolor[HTML]{FFAB6B}{4.84} & \cellcolor[HTML]{FFA25B}{5.39} & \cellcolor[HTML]{FF9C50}{5.75} & \cellcolor[HTML]{FF9E53}{5.65} & \cellcolor[HTML]{FF9848}{6} \\
 & CodeLlama-7b & \cellcolor[HTML]{FFE0C8}{1.78} & \cellcolor[HTML]{FF9F56}{5.55} & \cellcolor[HTML]{FFA45E}{5.29} & \cellcolor[HTML]{FFA35D}{5.32} & \cellcolor[HTML]{FFA25B}{5.38} & \cellcolor[HTML]{FFA25B}{5.37} & \cellcolor[HTML]{FF9848}{6}.07 & \cellcolor[HTML]{FF9848}{6}.19 & \cellcolor[HTML]{FF9848}{6}.12 \\
\multirow{-7}{*}{Java} & CodeLlama-13b & \cellcolor[HTML]{FFD1AD}{2.68} & \cellcolor[HTML]{FFB278}{4.44} & \cellcolor[HTML]{FFCBA3}{3}.76 & \cellcolor[HTML]{FFCBA3}{3}.97 & \cellcolor[HTML]{FFBA85}{4.01} & \cellcolor[HTML]{FFB881}{4.12} & \cellcolor[HTML]{FFB57C}{4.28} & \cellcolor[HTML]{FFA967}{4.98} & \cellcolor[HTML]{FFAD6E}{4.74} \\ \midrule
 & InCoder-1b & \cellcolor[HTML]{FFA763}{5.11} & \cellcolor[HTML]{FFCBA3}{3}.97 & \cellcolor[HTML]{FFB881}{4.12} & \cellcolor[HTML]{FFB57C}{4.28} & \cellcolor[HTML]{FFB47B}{4.32} & \cellcolor[HTML]{FFB57C}{4.3} & \cellcolor[HTML]{FFB276}{4.48} & \cellcolor[HTML]{FFCBA3}{3}.98 & \cellcolor[HTML]{FFB77F}{4.19} \\
 & CodeGen-2b-nl & \cellcolor[HTML]{FFD2B0}{2.58} & \cellcolor[HTML]{FFCBA3}{3}.23 & \cellcolor[HTML]{FFCBA3}{3}.24 & \cellcolor[HTML]{FFCBA3}{3}.28 & \cellcolor[HTML]{FFCBA3}{3}.26 & \cellcolor[HTML]{FFCBA3}{3}.27 & \cellcolor[HTML]{FFCBA3}{3}.21 & \cellcolor[HTML]{FFCBA3}{3}.24 & \cellcolor[HTML]{FFCBA3}{3}.44 \\
 & CodeGen-6b-nl & \cellcolor[HTML]{FFCBA3}{3}.4 & \cellcolor[HTML]{FFB77F}{4.19} & \cellcolor[HTML]{FFAE6F}{4.71} & \cellcolor[HTML]{FFB278}{4.43} & \cellcolor[HTML]{FFB57C}{4.3}4 & \cellcolor[HTML]{FFB77F}{4.19} & \cellcolor[HTML]{FFB67E}{4.24} & \cellcolor[HTML]{FFB57C}{4.3}5 & \cellcolor[HTML]{FFB57C}{4.3}5 \\
 & Llama-2-7b & \cellcolor[HTML]{FFCBA3}{3}.53 & \cellcolor[HTML]{FFB378}{4.41} & \cellcolor[HTML]{FFB073}{4.59} & \cellcolor[HTML]{FF9F55}{5.58} & \cellcolor[HTML]{FF9848}{6}.19 & \cellcolor[HTML]{FF9848}{6}.37 & \cellcolor[HTML]{FF9848}{6}.3 & \cellcolor[HTML]{FF9848}{6}.05 & \cellcolor[HTML]{FF9848}{6}.33 \\
 & Llama-2-13b & \cellcolor[HTML]{FFB378}{4.41} & \cellcolor[HTML]{FFA763}{5.12} & \cellcolor[HTML]{FF9848}{6}.15 & \cellcolor[HTML]{FF9848}{6}.75 & \cellcolor[HTML]{FF8325}{7.17} & \cellcolor[HTML]{FF801F}{7.36} & \cellcolor[HTML]{FF7D1A}{7.52} & \cellcolor[HTML]{FF8221}{7.28} & \cellcolor[HTML]{FF7C17}{7.63} \\
 & CodeLlama-7b & \cellcolor[HTML]{FFE5D1}{1.51} & \cellcolor[HTML]{FF9848}{6}.19 & \cellcolor[HTML]{FF8221}{7.28} & \cellcolor[HTML]{FF7912}{7.79} & \cellcolor[HTML]{FF6F00}{8.36} & \cellcolor[HTML]{FF6F00}{8.39} & \cellcolor[HTML]{FF7307}{8.13} & \cellcolor[HTML]{FF7206}{8.18} & \cellcolor[HTML]{FF7409}{8.07} \\
\multirow{-7}{*}{Python} & CodeLlama-13b & \cellcolor[HTML]{FFCBA3}{3}.63 & \cellcolor[HTML]{FFA45F}{5.26} & \cellcolor[HTML]{FF9848}{6}.42 & \cellcolor[HTML]{FF8221}{7.28} & \cellcolor[HTML]{FF7A14}{7.7} & \cellcolor[HTML]{FF7C18}{7.59} & \cellcolor[HTML]{FF9848}{6}.95 & \cellcolor[HTML]{FF8121}{7.3} & \cellcolor[HTML]{FF9848}{6}.29 \\ \midrule
 & InCoder-1b & \cellcolor[HTML]{FFCBA3}{3}.78 & \cellcolor[HTML]{FFCBA3}{3}.75 & \cellcolor[HTML]{FFCBA3}{3}.9 & \cellcolor[HTML]{FFB984}{4.04} & \cellcolor[HTML]{FFCBA3}{3}.82 & \cellcolor[HTML]{FFCBA3}{3}.83 & \cellcolor[HTML]{FFCBA3}{3}.7 & \cellcolor[HTML]{FFCBA3}{3}.53 & \cellcolor[HTML]{FFCBA3}{3}.51 \\
 & CodeGen-2b-nl & \cellcolor[HTML]{FFDCC1}{2.02} & \cellcolor[HTML]{FFD0AC}{2.7} & \cellcolor[HTML]{FFD2AF}{2.61} & \cellcolor[HTML]{FFD4B3}{2.47} & \cellcolor[HTML]{FFD3B1}{2.55} & \cellcolor[HTML]{FFD2AF}{2.62} & \cellcolor[HTML]{FFD2B0}{2.59} & \cellcolor[HTML]{FFD4B4}{2.46} & \cellcolor[HTML]{FFD3B1}{2.54} \\
 & CodeGen-6b-nl & \cellcolor[HTML]{FFD9BB}{2.21} & \cellcolor[HTML]{FFCBA3}{3}.28 & \cellcolor[HTML]{FFCBA3}{3}.57 & \cellcolor[HTML]{FFCBA3}{3}.76 & \cellcolor[HTML]{FFCBA3}{3}.71 & \cellcolor[HTML]{FFCBA3}{3}.48 & \cellcolor[HTML]{FFCBA3}{3}.52 & \cellcolor[HTML]{FFCBA3}{3}.48 & \cellcolor[HTML]{FFCBA3}{3}.45 \\
 & Llama-2-7b & \cellcolor[HTML]{FFCBA3}{3}.97 & \cellcolor[HTML]{FFB57D}{4.26} & \cellcolor[HTML]{FFB074}{4.56} & \cellcolor[HTML]{FFA865}{5.04} & \cellcolor[HTML]{FFA15A}{5.42} & \cellcolor[HTML]{FF9B4E}{5.8} & \cellcolor[HTML]{FF9F55}{5.57} & \cellcolor[HTML]{FFA561}{5.19} & \cellcolor[HTML]{FFA560}{5.22} \\
 & Llama-2-13b & \cellcolor[HTML]{FFCBA3}{3}.11 & \cellcolor[HTML]{FFAC6D}{4.8} & \cellcolor[HTML]{FFA25B}{5.37} & \cellcolor[HTML]{FF9848}{6}.05 & \cellcolor[HTML]{FF9849}{5.97} & \cellcolor[HTML]{FF9849}{5.98} & \cellcolor[HTML]{FF9848}{6}.18 & \cellcolor[HTML]{FF9848}{6}.01 & \cellcolor[HTML]{FF9848}{6}.11 \\
 & CodeLlama-7b & \cellcolor[HTML]{FFE4CF}{1.57} & \cellcolor[HTML]{FF9848}{6}.32 & \cellcolor[HTML]{FF9848}{6}.68 & \cellcolor[HTML]{FF9848}{6}.84 & \cellcolor[HTML]{FF8528}{7.05} & \cellcolor[HTML]{FF9848}{6}.74 & \cellcolor[HTML]{FF9848}{6}.71 & \cellcolor[HTML]{FF9848}{6}.89 & \cellcolor[HTML]{FF9848}{6}.91 \\
\multirow{-7}{*}{C\#} & CodeLlama-13b & \cellcolor[HTML]{FFD4B4}{2.46} & \cellcolor[HTML]{FFB176}{4.49} & \cellcolor[HTML]{FFA15A}{5.42} & \cellcolor[HTML]{FF9F56}{5.54} & \cellcolor[HTML]{FF9848}{6}.65 & \cellcolor[HTML]{FF9848}{6}.54 & \cellcolor[HTML]{FF9848}{6}.28 & \cellcolor[HTML]{FF9848}{6}.51 & \cellcolor[HTML]{FF9848}{6}.21 \\ \midrule
 & InCoder-1b & \cellcolor[HTML]{FFB57C}{4.3}1 & \cellcolor[HTML]{FFCBA3}{3}.8 & \cellcolor[HTML]{FFCBA3}{3}.7 & \cellcolor[HTML]{FFCBA3}{3}.95 & \cellcolor[HTML]{FFCBA3}{3}.73 & \cellcolor[HTML]{FFB983}{4.06} & \cellcolor[HTML]{FFCBA3}{3}.61 & \cellcolor[HTML]{FFCBA3}{3}.86 & \cellcolor[HTML]{FFCBA3}{3}.99 \\
 & CodeGen-2b-nl & \cellcolor[HTML]{FFD4B3}{2.5} & \cellcolor[HTML]{FFCBA3}{3}.3 & \cellcolor[HTML]{FFCBA3}{3}.39 & \cellcolor[HTML]{FFCBA3}{3}.38 & \cellcolor[HTML]{FFCBA3}{3}.39 & \cellcolor[HTML]{FFCBA3}{3}.29 & \cellcolor[HTML]{FFCBA3}{3}.4 & \cellcolor[HTML]{FFCBA3}{3}.18 & \cellcolor[HTML]{FFCBA3}{3}.15 \\
 & CodeGen-6b-nl & \cellcolor[HTML]{FFCCA5}{2.96} & \cellcolor[HTML]{FFB780}{4.16} & \cellcolor[HTML]{FFB67E}{4.23} & \cellcolor[HTML]{FFB57C}{4.3}8 & \cellcolor[HTML]{FFB276}{4.48} & \cellcolor[HTML]{FFB57C}{4.3}6 & \cellcolor[HTML]{FFB57C}{4.3}3 & \cellcolor[HTML]{FFB77F}{4.19} & \cellcolor[HTML]{FFB780}{4.16} \\
 & Llama-2-7b & \cellcolor[HTML]{FFCCA5}{2.93} & \cellcolor[HTML]{FFB176}{4.49} & \cellcolor[HTML]{FFB57C}{4.3}8 & \cellcolor[HTML]{FFB379}{4.4} & \cellcolor[HTML]{FFAC6D}{4.8}1 & \cellcolor[HTML]{FFAC6D}{4.8}8 & \cellcolor[HTML]{FFA966}{5.01} & \cellcolor[HTML]{FFA25A}{5.41} & \cellcolor[HTML]{FFA15A}{5.42} \\
 & Llama-2-13b & \cellcolor[HTML]{FFCBA3}{3}.43 & \cellcolor[HTML]{FFB47B}{4.32} & \cellcolor[HTML]{FFA561}{5.19} & \cellcolor[HTML]{FFA35D}{5.32} & \cellcolor[HTML]{FFA057}{5.52} & \cellcolor[HTML]{FF994B}{5.92} & \cellcolor[HTML]{FF9B4E}{5.8}8 & \cellcolor[HTML]{FF9848}{6}.19 & \cellcolor[HTML]{FF9848}{6}.05 \\
 & CodeLlama-7b & \cellcolor[HTML]{FFEADB}{1.18} & \cellcolor[HTML]{FF9B4F}{5.79} & \cellcolor[HTML]{FF9848}{6}.18 & \cellcolor[HTML]{FF9848}{6}.16 & \cellcolor[HTML]{FF9848}{6}.26 & \cellcolor[HTML]{FF9848}{6}.22 & \cellcolor[HTML]{FF9848}{6}.3 & \cellcolor[HTML]{FF9848}{6}.7 & \cellcolor[HTML]{FF9848}{6}.51 \\
\multirow{-7}{*}{C++} & CodeLlama-13b & \cellcolor[HTML]{FFD2AF}{2.6} & \cellcolor[HTML]{FFA45E}{5.29} & \cellcolor[HTML]{FF9D52}{5.68} & \cellcolor[HTML]{FF9F55}{5.58} & \cellcolor[HTML]{FF9B4E}{5.8}9 & \cellcolor[HTML]{FF9B4E}{5.8}2 & \cellcolor[HTML]{FF9B4E}{5.8}2 & \cellcolor[HTML]{FF9B4E}{5.8}7 & \cellcolor[HTML]{FF994B}{5.91} \\ \midrule
 & InCoder-1b & \cellcolor[HTML]{FFB881}{4.12} & \cellcolor[HTML]{FFCBA3}{3}.37 & \cellcolor[HTML]{FFCBA3}{3}.35 & \cellcolor[HTML]{FFCBA3}{3}.25 & \cellcolor[HTML]{FFCBA3}{3}.2 & \cellcolor[HTML]{FFCBA3}{3}.37 & \cellcolor[HTML]{FFCBA3}{3}.32 & \cellcolor[HTML]{FFCBA3}{3}.53 & \cellcolor[HTML]{FFB983}{4.07} \\
 & CodeGen-2b-nl & \cellcolor[HTML]{FFD0AC}{2.7}6 & \cellcolor[HTML]{FFCBA3}{3}.2 & \cellcolor[HTML]{FFCBA3}{3}.23 & \cellcolor[HTML]{FFCEA9}{2.8}1 & \cellcolor[HTML]{FFD0AC}{2.7}2 & \cellcolor[HTML]{FFD2AF}{2.6}9 & \cellcolor[HTML]{FFD1AD}{2.68} & \cellcolor[HTML]{FFD0AC}{2.7}5 & \cellcolor[HTML]{FFCEA9}{2.8}6 \\
 & CodeGen-6b-nl & \cellcolor[HTML]{FFCBA3}{3}.11 & \cellcolor[HTML]{FFCBA3}{3}.23 & \cellcolor[HTML]{FFCBA3}{3}.52 & \cellcolor[HTML]{FFCBA3}{3}.95 & \cellcolor[HTML]{FFCBA3}{3}.94 & \cellcolor[HTML]{FFCBA3}{3}.89 & \cellcolor[HTML]{FFCBA3}{3}.62 & \cellcolor[HTML]{FFCBA3}{3}.71 & \cellcolor[HTML]{FFCBA3}{3}.38 \\
 & Llama-2-7b & \cellcolor[HTML]{FFCFAA}{2.79} & \cellcolor[HTML]{FFCBA3}{3}.72 & \cellcolor[HTML]{FFCBA3}{3}.63 & \cellcolor[HTML]{FFB882}{4.1} & \cellcolor[HTML]{FFBA85}{4.01} & \cellcolor[HTML]{FFAA6A}{4.9} & \cellcolor[HTML]{FF9F55}{5.59} & \cellcolor[HTML]{FF9F56}{5.55} & \cellcolor[HTML]{FF9848}{6}.11 \\
 & Llama-2-13b & \cellcolor[HTML]{FFCBA3}{3}.3 & \cellcolor[HTML]{FFB074}{4.56} & \cellcolor[HTML]{FFA55F}{5.24} & \cellcolor[HTML]{FF9848}{6}.81 & \cellcolor[HTML]{FF9849}{5.97} & \cellcolor[HTML]{FF8528}{7.05} & \cellcolor[HTML]{FF7B15}{7.67} & \cellcolor[HTML]{FF770F}{7.87} & \cellcolor[HTML]{FF7A14}{7.7}7 \\
 & CodeLlama-7b & \cellcolor[HTML]{FFE6D3}{1.43} & \cellcolor[HTML]{FFAC6D}{4.8}2 & \cellcolor[HTML]{FF9848}{6}.17 & \cellcolor[HTML]{FF8121}{7.3}8 & \cellcolor[HTML]{FF8426}{7.12} & \cellcolor[HTML]{FF7912}{7.79} & \cellcolor[HTML]{FF7811}{7.81} & \cellcolor[HTML]{FF7B15}{7.67} & \cellcolor[HTML]{FF7810}{7.86} \\
\multirow{-7}{*}{Javascript} & CodeLlama-13b & \cellcolor[HTML]{FFD8BA}{2.24} & \cellcolor[HTML]{FFA45E}{5.29} & \cellcolor[HTML]{FFAC6D}{4.8}5 & \cellcolor[HTML]{FFAA6A}{4.9}4 & \cellcolor[HTML]{FFA661}{5.18} & \cellcolor[HTML]{FF9F55}{5.57} & \cellcolor[HTML]{FF9848}{6}.24 & \cellcolor[HTML]{FF9848}{6}.84 & \cellcolor[HTML]{FF9848}{6}.6 \\ \bottomrule
\end{tabular}
}
\end{table*}

\begin{table*}[!h]
\caption{Performance of LLM-ICL on CMG with code as input.}
\centering
\label{tab:append-cmg-code}
\resizebox{\textwidth}{!}{
\begin{tabular}{clccccccccc}
\toprule
\multicolumn{1}{l}{Lang} & Model & 0shot & 1shot & 2shot & 3shot & 4shot & 5shot & 6shot & 7shot & 8shot \\ \midrule
 & InCoder-1b & \cellcolor[HTML]{E7F2FF}{3.08} & \cellcolor[HTML]{FAFDFC}{3} & \cellcolor[HTML]{F6FAFF}{2.62} & \cellcolor[HTML]{F7FAFF}{2.74} & \cellcolor[HTML]{FAF9F9}{2.66} & \cellcolor[HTML]{FAFCFF}{2.75} & \cellcolor[HTML]{FFFFFF}{2.79} & \cellcolor[HTML]{F4F8FF}{2.76} & \cellcolor[HTML]{FAF9F9}{2.66} \\
 & CodeGen-2b-nl & \cellcolor[HTML]{FEFEFF}{1.88} & \cellcolor[HTML]{F4F9FF}{1.98} & \cellcolor[HTML]{ECF4FF}{2} & \cellcolor[HTML]{ECF4FF}{2}.11 & \cellcolor[HTML]{ECF4FF}{2}.28 & \cellcolor[HTML]{ECF4FF}{2}.29 & \cellcolor[HTML]{ECF4FF}{2}.29 & \cellcolor[HTML]{ECF4FF}{2}.36 & \cellcolor[HTML]{ECF4FF}{2}.33 \\
 & CodeGen-6b-nl & \cellcolor[HTML]{ECF4FF}{2}.46 & \cellcolor[HTML]{ECF4FF}{2}.58 & \cellcolor[HTML]{ECF4FF}{2}.94 & \cellcolor[HTML]{FAFDFC}{3}.24 & \cellcolor[HTML]{FAFDFC}{3}.19 & \cellcolor[HTML]{FAFDFC}{3}.52 & \cellcolor[HTML]{FAFDFC}{3}.43 & \cellcolor[HTML]{FAFDFC}{3}.39 & \cellcolor[HTML]{FAFDFC}{3}.43 \\
 & Llama-2-7b & \cellcolor[HTML]{CDE3FF}{1.13} & \cellcolor[HTML]{FAFDFC}{3}.28 & \cellcolor[HTML]{FAFDFC}{3}.51 & \cellcolor[HTML]{FAFDFC}{3}.88 & \cellcolor[HTML]{FAFDFC}{3}.8 & \cellcolor[HTML]{FAFDFC}{3}.63 & \cellcolor[HTML]{FAFDFC}{3}.54 & \cellcolor[HTML]{FAFDFC}{3}.42 & \cellcolor[HTML]{FAFDFC}{3}.54 \\
 & Llama-2-13b & \cellcolor[HTML]{E0EDFF}{1.74} & \cellcolor[HTML]{FAFDFC}{3}.35 & \cellcolor[HTML]{FAFDFC}{3}.91 & \cellcolor[HTML]{F0F6FF}{4.25} & \cellcolor[HTML]{E3EFFF}{4.23} & \cellcolor[HTML]{D9EAFF}{4.57} & \cellcolor[HTML]{D4E7FF}{4.82} & \cellcolor[HTML]{DDECFF}{4.91} & \cellcolor[HTML]{D7E8FF}{5.13} \\
 & CodeLlama-7b & \cellcolor[HTML]{FAFDFC}{1.81} & \cellcolor[HTML]{FAF0EE}{5.78} & \cellcolor[HTML]{F9FBFF}{5.17} & \cellcolor[HTML]{FADCD8}{5.86} & \cellcolor[HTML]{FAE8E6}{5.73} & \cellcolor[HTML]{FAE8E6}{5.73} & \cellcolor[HTML]{F0F6FF}{5.74} & \cellcolor[HTML]{DFEDFF}{5.49} & \cellcolor[HTML]{E1EEFF}{5.47} \\
\multirow{-7}{*}{Java} & CodeLlama-13b & \cellcolor[HTML]{D2E5FF}{1.69} & \cellcolor[HTML]{FAFDFC}{4.47} & \cellcolor[HTML]{FAEAE8}{4.08} & \cellcolor[HTML]{FAF4F3}{4.14} & \cellcolor[HTML]{FAE9E7}{4.34} & \cellcolor[HTML]{FAE7E4}{4.49} & \cellcolor[HTML]{FACEC9}{5.04} & \cellcolor[HTML]{EEF5FF}{4.61} & \cellcolor[HTML]{EAF3FF}{4.28} \\ \midrule
 & InCoder-1b & \cellcolor[HTML]{E5F0FF}{4.54} & \cellcolor[HTML]{FDFDFF}{3.93} & \cellcolor[HTML]{FAF7F7}{4.23} & \cellcolor[HTML]{F2F8FF}{4.01} & \cellcolor[HTML]{EBF4FF}{3.9} & \cellcolor[HTML]{EBF4FF}{3.88} & \cellcolor[HTML]{EBF4FF}{3.9}4 & \cellcolor[HTML]{FAF9F8}{4.07} & \cellcolor[HTML]{EBF4FF}{3.9}2 \\
 & CodeGen-2b-nl & \cellcolor[HTML]{FAE4E2}{2.99} & \cellcolor[HTML]{F8FBFF}{3.09} & \cellcolor[HTML]{F4F9FF}{3.01} & \cellcolor[HTML]{FEFEFF}{3.26} & \cellcolor[HTML]{FAFDFD}{3.28} & \cellcolor[HTML]{FAF1F0}{3.48} & \cellcolor[HTML]{FAE9E7}{3.55} & \cellcolor[HTML]{FAE9E7}{3.57} & \cellcolor[HTML]{FCFDFF}{3.38} \\
 & CodeGen-6b-nl & \cellcolor[HTML]{FAD3CF}{4.08} & \cellcolor[HTML]{FAF4F3}{4.35} & \cellcolor[HTML]{EFF6FF}{4.36} & \cellcolor[HTML]{FAEDEB}{4.71} & \cellcolor[HTML]{FAEDEB}{4.62} & \cellcolor[HTML]{FAEDEB}{4.47} & \cellcolor[HTML]{FAEBE9}{4.55} & \cellcolor[HTML]{FAE1DE}{4.81} & \cellcolor[HTML]{FAE4E1}{4.77} \\
 & Llama-2-7b & \cellcolor[HTML]{BFDBFF}{2.12} & \cellcolor[HTML]{FAEDEB}{4.68} & \cellcolor[HTML]{FAE9E7}{4.93} & \cellcolor[HTML]{EDF5FF}{5.19} & \cellcolor[HTML]{D3E6FF}{5.24} & \cellcolor[HTML]{CEE4FF}{5.31} & \cellcolor[HTML]{C6DFFF}{5.05} & \cellcolor[HTML]{C7E0FF}{4.83} & \cellcolor[HTML]{B2D4FF}{4.64} \\
 & Llama-2-13b & \cellcolor[HTML]{D3E6FF}{3.45} & \cellcolor[HTML]{E7F1FF}{4.6} & \cellcolor[HTML]{E4F0FF}{5.56} & \cellcolor[HTML]{F5F9FF}{6.53} & \cellcolor[HTML]{EAF3FF}{6.71} & \cellcolor[HTML]{E3EFFF}{6.75} & \cellcolor[HTML]{CCE2FF}{6.41} & \cellcolor[HTML]{B5D6FF}{5.67} & \cellcolor[HTML]{A1CAFF}{5.57} \\
 & CodeLlama-7b & \cellcolor[HTML]{FEFEFF}{1.5} & \cellcolor[HTML]{E4F0FF}{5.6} & \cellcolor[HTML]{F3F8FF}{7.03} & \cellcolor[HTML]{EFF6FF}{7.45} & \cellcolor[HTML]{F1F7FF}{8.07} & \cellcolor[HTML]{FAF5F4}{8.54} & \cellcolor[HTML]{FAFCFF}{8.02} & \cellcolor[HTML]{D0E5FF}{7.16} & \cellcolor[HTML]{BCD9FF}{6.6} \\
\multirow{-7}{*}{Python} & CodeLlama-13b & \cellcolor[HTML]{C3DDFF}{2.31} & \cellcolor[HTML]{EEF5FF}{4.9} & \cellcolor[HTML]{FDFEFF}{6.39} & \cellcolor[HTML]{E3EFFF}{6.75} & \cellcolor[HTML]{CCE2FF}{6.59} & \cellcolor[HTML]{F0F7FF}{7.28} & \cellcolor[HTML]{BCD9FF}{6.6}2 & \cellcolor[HTML]{E4F0FF}{5.56} & \cellcolor[HTML]{DCEBFF}{5.52} \\ \midrule
 & InCoder-1b & \cellcolor[HTML]{DEECFF}{3.06} & \cellcolor[HTML]{F9FBFF}{3.62} & \cellcolor[HTML]{F1F7FF}{3.61} & \cellcolor[HTML]{F2F8FF}{3.77} & \cellcolor[HTML]{FAFEFE}{3.83} & \cellcolor[HTML]{EBF4FF}{3.41} & \cellcolor[HTML]{FAFBFA}{3.76} & \cellcolor[HTML]{F8FBFF}{3.38} & \cellcolor[HTML]{FAF1F0}{3.72} \\
 & CodeGen-2b-nl & \cellcolor[HTML]{FAFAFA}{2.09} & \cellcolor[HTML]{FAFCFF}{2.59} & \cellcolor[HTML]{FAF7F6}{2.73} & \cellcolor[HTML]{FAE4E2}{2.88} & \cellcolor[HTML]{FADFDC}{3.04} & \cellcolor[HTML]{FAE7E4}{2.99} & \cellcolor[HTML]{FAE0DD}{3.07} & \cellcolor[HTML]{FADEDB}{2.97} & \cellcolor[HTML]{FAECEA}{2.83} \\
 & CodeGen-6b-nl & \cellcolor[HTML]{FAEBE9}{2.51} & \cellcolor[HTML]{F9FBFF}{3.15} & \cellcolor[HTML]{FAFCFF}{3.48} & \cellcolor[HTML]{F9FBFF}{3.62} & \cellcolor[HTML]{FEFEFF}{3.7} & \cellcolor[HTML]{FEFEFF}{3.7}8 & \cellcolor[HTML]{FAEAE8}{3.84} & \cellcolor[HTML]{FAF1F0}{3.69} & \cellcolor[HTML]{FAF0EF}{3.67} \\
 & Llama-2-7b & \cellcolor[HTML]{89BDFF}{1.37} & \cellcolor[HTML]{FAFDFD}{4.28} & \cellcolor[HTML]{FAE3E0}{4.99} & \cellcolor[HTML]{FAEDEB}{5.31} & \cellcolor[HTML]{F1F7FF}{5.13} & \cellcolor[HTML]{E5F0FF}{5.23} & \cellcolor[HTML]{E4F0FF}{4.98} & \cellcolor[HTML]{EFF6FF}{4.85} & \cellcolor[HTML]{D1E5FF}{4.22} \\
 & Llama-2-13b & \cellcolor[HTML]{CFE4FF}{2.06} & \cellcolor[HTML]{FBFDFF}{4.73} & \cellcolor[HTML]{FAF7F7}{5.48} & \cellcolor[HTML]{F1F7FF}{5.76} & \cellcolor[HTML]{EDF5FF}{5.59} & \cellcolor[HTML]{FAF9F8}{6.07} & \cellcolor[HTML]{EFF6FF}{5.84} & \cellcolor[HTML]{EDF5FF}{5.63} & \cellcolor[HTML]{D1E5FF}{5.1} \\
 & CodeLlama-7b & \cellcolor[HTML]{FAFBFA}{1.63} & \cellcolor[HTML]{F5F9FF}{6.12} & \cellcolor[HTML]{FAE8E5}{7.04} & \cellcolor[HTML]{FAD8D4}{7.44} & \cellcolor[HTML]{FAF9F9}{7.13} & \cellcolor[HTML]{FAEDEB}{7.01} & \cellcolor[HTML]{FAFCFF}{6.62} & \cellcolor[HTML]{F4F9FF}{6.66} & \cellcolor[HTML]{D3E6FF}{5.95} \\
\multirow{-7}{*}{C\#} & CodeLlama-13b & \cellcolor[HTML]{CEE4FF}{1.4} & \cellcolor[HTML]{F7FAFF}{4.33} & \cellcolor[HTML]{EFF6FF}{5.08} & \cellcolor[HTML]{FAEAE8}{5.86} & \cellcolor[HTML]{F8FBFF}{6.51} & \cellcolor[HTML]{FAFCFF}{6.62} & \cellcolor[HTML]{FAF6F5}{6.41} & \cellcolor[HTML]{DDECFF}{5.78} & \cellcolor[HTML]{BAD8FF}{4.69} \\ \midrule
 & InCoder-1b & \cellcolor[HTML]{DCEBFF}{3.55} & \cellcolor[HTML]{F2F8FF}{3.53} & \cellcolor[HTML]{F0F6FF}{3.37} & \cellcolor[HTML]{E8F2FF}{3.45} & \cellcolor[HTML]{FAFDFC}{3.76} & \cellcolor[HTML]{E6F1FF}{3.52} & \cellcolor[HTML]{FAF4F3}{3.77} & \cellcolor[HTML]{FDFEFF}{3.83} & \cellcolor[HTML]{FDFDFF}{3.95} \\
 & CodeGen-2b-nl & \cellcolor[HTML]{FAF2F1}{2.69} & \cellcolor[HTML]{F1F7FF}{3} & \cellcolor[HTML]{F1F7FF}{3}.18 & \cellcolor[HTML]{F1F7FF}{3}.16 & \cellcolor[HTML]{F1F7FF}{3}.21 & \cellcolor[HTML]{F1F7FF}{3}.36 & \cellcolor[HTML]{F1F7FF}{3}.4 & \cellcolor[HTML]{E6F1FF}{3.52} & \cellcolor[HTML]{F1F7FF}{3}.43 \\
 & CodeGen-6b-nl & \cellcolor[HTML]{FDFEFF}{2.93} & \cellcolor[HTML]{F1F7FF}{3}.75 & \cellcolor[HTML]{F1F7FF}{3}.8 & \cellcolor[HTML]{F1F7FF}{3}.81 & \cellcolor[HTML]{EBF4FF}{4.06} & \cellcolor[HTML]{F4F9FF}{4.13} & \cellcolor[HTML]{F2F7FF}{4.05} & \cellcolor[HTML]{FFFFFF}{4.19} & \cellcolor[HTML]{FAF3F2}{4.34} \\
 & Llama-2-7b & \cellcolor[HTML]{D5E7FF}{2.01} & \cellcolor[HTML]{FBFCFF}{4.41} & \cellcolor[HTML]{F8FBFF}{4.23} & \cellcolor[HTML]{FADFDC}{4.89} & \cellcolor[HTML]{D1E5FF}{5.1}5 & \cellcolor[HTML]{D1E5FF}{5.1}5 & \cellcolor[HTML]{F7FAFF}{4.84} & \cellcolor[HTML]{D7E8FF}{4.54} & \cellcolor[HTML]{D7E8FF}{4.55} \\
 & Llama-2-13b & \cellcolor[HTML]{DAEAFF}{2.62} & \cellcolor[HTML]{FEFEFF}{4.3} & \cellcolor[HTML]{F6FAFF}{5.01} & \cellcolor[HTML]{FAEBE9}{5.63} & \cellcolor[HTML]{FAFAFA}{5.59} & \cellcolor[HTML]{F7FAFF}{5.76} & \cellcolor[HTML]{F8FBFF}{5.74} & \cellcolor[HTML]{D5E7FF}{5.27} & \cellcolor[HTML]{E7F1FF}{5.53} \\
 & CodeLlama-7b & \cellcolor[HTML]{FAFBFA}{1.24} & \cellcolor[HTML]{D1E5FF}{5.1}2 & \cellcolor[HTML]{F8FBFF}{5.74} & \cellcolor[HTML]{FCFDFF}{6.11} & \cellcolor[HTML]{FAF6F5}{6.39} & \cellcolor[HTML]{FAF0EE}{6.45} & \cellcolor[HTML]{FEFEFF}{6.29} & \cellcolor[HTML]{F8FBFF}{5.74} & \cellcolor[HTML]{E6F1FF}{5.98} \\
\multirow{-7}{*}{C++} & CodeLlama-13b & \cellcolor[HTML]{E7F2FF}{2.09} & \cellcolor[HTML]{F3F8FF}{5.03} & \cellcolor[HTML]{F5F9FF}{5.46} & \cellcolor[HTML]{FAC9C3}{6.42} & \cellcolor[HTML]{FAD9D6}{6.47} & \cellcolor[HTML]{FAE8E5}{6.18} & \cellcolor[HTML]{FAEBE9}{6.13} & \cellcolor[HTML]{F6FAFF}{5.69} & \cellcolor[HTML]{F1F7FF}{5.61} \\ \midrule
 & InCoder-1b & \cellcolor[HTML]{F1F7FF}{3}.21 & \cellcolor[HTML]{F1F7FF}{3}.06 & \cellcolor[HTML]{F1F7FF}{3}.44 & \cellcolor[HTML]{F1F7FF}{3}.34 & \cellcolor[HTML]{F1F7FF}{3}.21 & \cellcolor[HTML]{F1F7FF}{3}.47 & \cellcolor[HTML]{F1F7FF}{3}.82 & \cellcolor[HTML]{FAF0EF}{3.67} & \cellcolor[HTML]{FDFEFF}{3.83} \\
 & CodeGen-2b-nl & \cellcolor[HTML]{F1F7FF}{3}.09 & \cellcolor[HTML]{F1F7FF}{3}.42 & \cellcolor[HTML]{F1F7FF}{3}.27 & \cellcolor[HTML]{F1F7FF}{3}.16 & \cellcolor[HTML]{F1F7FF}{3}.19 & \cellcolor[HTML]{DCEBFF}{3.55} & \cellcolor[HTML]{F1F7FF}{3}.28 & \cellcolor[HTML]{F1F7FF}{3}.35 & \cellcolor[HTML]{F1F7FF}{3}.49 \\
 & CodeGen-6b-nl & \cellcolor[HTML]{F1F7FF}{3}.29 & \cellcolor[HTML]{F1F7FF}{3}.73 & \cellcolor[HTML]{F1F7FF}{3}.93 & \cellcolor[HTML]{F1F7FF}{3}.78 & \cellcolor[HTML]{F1F7FF}{3}.93 & \cellcolor[HTML]{FAF4F3}{3.77} & \cellcolor[HTML]{F1F7FF}{3}.89 & \cellcolor[HTML]{F1F7FF}{3}.84 & \cellcolor[HTML]{FDFEFF}{3.83} \\
 & Llama-2-7b & \cellcolor[HTML]{BBD9FF}{1.3} & \cellcolor[HTML]{F1F7FF}{3}.74 & \cellcolor[HTML]{FDFDFF}{3.95} & \cellcolor[HTML]{FAF7F7}{4.21} & \cellcolor[HTML]{FAF6F5}{4.15} & \cellcolor[HTML]{FBFDFF}{4.83} & \cellcolor[HTML]{C3DDFF}{4.28} & \cellcolor[HTML]{D2E6FF}{4.58} & \cellcolor[HTML]{AACFFF}{4.25} \\
 & Llama-2-13b & \cellcolor[HTML]{CDE3FF}{2.22} & \cellcolor[HTML]{F7FAFF}{4.4} & \cellcolor[HTML]{F3F8FF}{4.99} & \cellcolor[HTML]{CEE4FF}{5.75} & \cellcolor[HTML]{FAF0EE}{6.2} & \cellcolor[HTML]{FAFCFF}{6.94} & \cellcolor[HTML]{C1DCFF}{6.31} & \cellcolor[HTML]{BCD9FF}{6.41} & \cellcolor[HTML]{B2D4FF}{6.08} \\
 & CodeLlama-7b & \cellcolor[HTML]{FAF9F8}{1.52} & \cellcolor[HTML]{F1F7FF}{3}.98 & \cellcolor[HTML]{FAC3BD}{7.1} & \cellcolor[HTML]{FADFDB}{7.88} & \cellcolor[HTML]{FABCB5}{8.17} & \cellcolor[HTML]{FAC0B9}{8.77} & \cellcolor[HTML]{FAF1F0}{8.02} & \cellcolor[HTML]{FAF5F4}{7.82} & \cellcolor[HTML]{F7FAFF}{7.7} \\
\multirow{-7}{*}{Javascript} & CodeLlama-13b & \cellcolor[HTML]{BBD9FF}{1.3}9 & \cellcolor[HTML]{C8E0FF}{4.09} & \cellcolor[HTML]{FEFEFF}{4.3}8 & \cellcolor[HTML]{FAF8F7}{5.04} & \cellcolor[HTML]{FAF2F0}{5.38} & \cellcolor[HTML]{FAD7D3}{6.19} & \cellcolor[HTML]{E6F1FF}{5.7} & \cellcolor[HTML]{BFDBFF}{5.45} & \cellcolor[HTML]{D3E6FF}{5.65} \\ \bottomrule
\end{tabular}
}
\end{table*}

\end{document}